\documentclass[12pt]{article}

\usepackage[reqno]{amsmath}
\usepackage{amsfonts,amssymb}
\usepackage[dvipdfm]{graphicx}
\usepackage{hyperref}

\def\rf#1{(\ref{eq:#1})}
\def\lab#1{\label{eq:#1}}
\def\nonu{\nonumber}
\def\br{\begin{eqnarray}}
\def\er{\end{eqnarray}}
\def\be{\begin{equation}}
\def\ee{\end{equation}}
\def\({\left(}
\def\){\right)}

\def\rlx{\relax\leavevmode}
\def\IR{\rlx\hbox{\rm I\kern-.18em R}}
\def\vp{\varphi}
\def\u2{\mid u\mid^2}

\def\IZ{\rlx\hbox{\sf Z\kern-.4em Z}}
\def\IR{\rlx\hbox{\rm I\kern-.18em R}}
\def\IC{\rlx\hbox{\,$\inbar\kern-.3em{\rm C}$}}

\input{epsf}

\usepackage{graphics}

\begin{document}

\begin{titlepage}
\vspace*{-1cm}

\vskip 1cm

\vspace{.2in}
\begin{center}
{\large\bf Vortices in the extended Skyrme-Faddeev model}
\end{center}

\vspace{.5cm}

\begin{center}
  L. A. Ferreira$^{a,}$\footnote{e-mail: {\tt laf@ifsc.usp.br}};
  J. J\"aykk\"a$^{b,}$\footnote{e-mail: {\tt juhaj@iki.fi}}; Nobuyuki
  Sawado$^{c,}$\footnote{e-mail: {\tt sawado@ph.noda.tus.ac.jp}}; Kouichi
  Toda$^{d,}$\footnote{e-mail: {\tt kouichi@yukawa.kyoto-u.ac.jp}};

\vspace{.5 in}
\small

\par \vskip .2in \noindent
$^{a}$ Instituto de F\'\i sica de S\~ao Carlos; IFSC/USP;\\
Universidade de S\~ao Paulo - USP \\ 
Caixa Postal 369, CEP 13560-970, S\~ao Carlos-SP, Brazil\\
$^{b}$ School of Mathematics, University of Leeds\\
LS2 9JT Leeds, United Kingdom\\
$^c$ Department of Physics, Tokyo University of Science,\\
 Noda, Chiba 278-8510, Japan\\
$^d$ Department of Mathematical Physics,
Toyama Prefectural University,\\ 
Kurokawa 5180, Imizu, Toyama, 
939-0398, Japan

\normalsize
\end{center}

\vspace{.2in}

\begin{abstract}

We construct analytical and numerical vortex solutions for an extended Skyrme-Faddeev model in a $(3+1)$ dimensional Minkowski space-time. The extension is obtained by adding to the Lagrangian  a quartic term, which is the square of the kinetic term, and a potential which breaks the $SO(3)$ symmetry down to $SO(2)$.  The construction makes use of an ansatz, invariant under the joint action of the internal $SO(2)$ and three commuting $U(1)$ subgroups of the Poincar\'e group, and which reduces the equations of motion to an ordinary differential equation for a profile function depending on the distance to the $x^3$-axis. The vortices have finite energy per unit length, and have waves propagating along them with the speed of light. The analytical vortices are obtained for special choice of potentials, and the numerical ones are constructed using the Successive Over
Relaxation method for more general potentials. The spectrum of solutions is analyzed in detail, specially its dependence upon special combinations of coupling constants.

\end{abstract}


\end{titlepage}

\section{Introduction}
\label{sec:intro}
\setcounter{equation}{0}

The so-called Skyrme-Faddeev model was introduced in the seventies \cite{sf} as a
ge\-ne\-ra\-li\-za\-tion to $(3+1)$ dimensions of the $O(3)$ non-linear sigma model in
$(2+1)$ dimensions \cite{bp}. The Skyrme term, quartic in derivatives of the field,
balances the quadratic kinetic term and according to Derrick's theorem, allows the
existence of stable solutions with non-trivial Hopf topological charges. Due to the highly
non-linear character of the model and the lack of symmetries, the first soliton solutions
were only constructed in the late nineties using numerical methods
\cite{glad,solfn,sutcliffe,hietarinta}. Since then the interest in the model has increased
considerably and it has found applications in many areas of physics due mainly to the
knotted character of the solutions \cite{babaev}. The numerical efforts in the
construction of the solutions have improved our understanding of the properties of the
model \cite{improve} and even the scattering of knotted solitons has been investigated
\cite{hietarinta-scatter}. One of the aspects of the model that has attracted considerable
attention has been its connection with gauge theories. Faddeev and Niemi have conjectured
that it might describe the low energy limit of the pure $SU(2)$ Yang-Mills theory
\cite{fn}. They based their argument on a decomposition of the physical degrees of freedom
of the $SU(2)$ connection, proposed in the eighties by Cho \cite{chofn}, and involving a
triplet of scalar fields ${\vec n}$ taking values on the sphere $S^2$ (${\vec n}^2=1$).
The conjecture, which is quite controversial \cite{wipf}, states that the low energy
effective action of the $SU(2)$ Yang-Mills theory is the Skyrme-Faddeev action, and the
knotted solitons would describe glueballs or even vacuum configurations. The fact that the
Skyrme-Faddeev model has an $O(3)$ symmetry, and so possesses Goldstone boson excitations,
is one of the many difficulties facing the conjecture, and some modifications of it were
in fact proposed \cite{newfaddeev}.  Any check of such type of conjectures is of course
very difficult to perform since it must involve non-perturbative calculations in the
strong coupling regime of the Yang-Mills theory. However, Gies \cite {gies} has calculated
the Wilsonian one loop effective action for the pure $SU(2)$ Yang-Mills theory assuming
Cho's decomposition, and found that the Skyrme-Faddeev action is indeed part of it, but
additional quartic terms in the derivatives of the triplet ${\vec n}$ are unavoidable.  In
fact, the first numerical Hopf solitons were first constructed for the Skyrme-Faddeev
model modified by a quartic term \cite{glad} which is the square of the kinetic
term. However, the soliton solutions in \cite{glad} were constructed for a sector of the
theory where the signs of the coupling constants disagree with those indicated by Gies'
calculations.  The addition of quartic terms has the drawback of making the Lagrangian
dependent on terms which are quartic in time derivatives and so the energy is not positive
definite. However, as a quantum field theory the Skyrme-Faddeev model is not
renormalizable by power counting and has to be considered as a low energy effective
theory. In addition, under the Wilsonian renormalization group flow the square of the
kinetic term is as unavoidable as the Skyrme quartic term. Therefore, it is quite natural
to investigate the properties of the Skyrme-Faddeev model with such modifications.

In this paper we consider an extended Skyrme-Faddeev model defined by the Lagrangian
\begin{eqnarray}
{\cal L} = M^2\, \partial_{\mu} {\vec n}\cdot\partial^{\mu} {\vec n}
 -\frac{1}{e^2} \, \(\partial_{\mu}{\vec n} \wedge 
\partial_{\nu}{\vec n}\)^2 + \frac{\beta}{2}\,
\left(\partial_{\mu} {\vec n}\cdot\partial^{\mu} {\vec n}\right)^2- V\(n_3\)
\lab{action}
\end{eqnarray}
where ${\vec n}$ is a triplet of real scalar fields taking values on the sphere $S^2$, 
$M$ is a coupling constant with dimension of $\({\rm length}\)^{-1}$, $e^2$ and $\beta$ are dimensionless coupling constants, and the
potential is a functional of the third component $n_3$ of the triplet ${\vec n}$. Note
that the potential breaks the $O(3)$ symmetry of the original Skyrme-Faddeev down to
$O(2)$, the group of rotations on the plane $n_1\, n_2$, and so eliminating two of the
three Goldstone boson degrees of freedom. In this paper the main role of potential is to
stabilize the vortex solutions.
 
The first exact vortex solutions for the theory \rf{action} were constructed in
\cite{vortexlaf} for the case where the potential vanishes, and by exploring the
integrability properties of a submodel of \rf{action}.  In order to describe those exact
vortex solutions it is better to perform the stereographic projection of the target space
$S^2$ onto the plane parameterized by the complex scalar field $u$ and related to ${\vec
  n}$ by
\begin{eqnarray}
{\vec n} = \(u+u^*,-i\(u-u^*\),\u2 -1\)/\(1+\u2\)
\lab{udef}
\end{eqnarray}
It was shown in \cite{vortexlaf} that the field configurations of the form 
\begin{eqnarray}
u\equiv u\(z,y\)\qquad \quad u^*\equiv u^*\( z^*,y\) \qquad \quad {\rm for} \qquad \quad
\beta\,e^2=1\qquad \quad V=0 \lab{exactclass}
\end{eqnarray}
are exact solutions of \rf{action}, where $z=x^1+i\,\varepsilon_1\,x^2$ and
$y=x^3-\varepsilon_2\,x^0$, with $\varepsilon_a=\pm 1$, $a=1,2$, and $x^{\mu}$,
$\mu=0,1,2,3$, are the Cartesian coordinates of the Minkowski space-time. Despite the fact
that \rf{exactclass} constitutes a very large class of solutions, no finite energy
solutions were found within it. If the dependence of the $u$ field upon the variable $y$
is in the form of phases like $e^{i\,k\,y}$, then one finds solutions with finite energy
per unit of length along the $x^3$-axis. The simplest solution is of the form
$u=z^n\,e^{i\,k\,y}$, with $n$ integer, and it corresponds to a vortex parallel to the
$x^3$-axis and with waves traveling along it with the speed of light. More general
solutions of the class \rf{exactclass} were constructed in \cite{newsf}, including
multi-vortices separated from each other and all parallel to the $x^3$-axis.  The ideas of
\cite{vortexlaf} were generalized to an extended Skyrme-Faddeev defined on the target
space $CP^N$, possessing $N-1$ complex scalar fields $u_i$, and the class of solutions
constructed is like \rf{exactclass}, where the fields $u_i$'s are arbitrary functions of
$z$ and $y$ \cite{CPNvortex}.  Note that the solutions \rf{exactclass} are not solutions
of the original Skyrme-Faddeev model, since that corresponds to $\beta=0$, and
\rf{exactclass} requires the condition $\beta\,e^2=1$. If one takes the limit
$\beta\rightarrow 0$ and $\frac{1}{e^2}\rightarrow 0$ with keeping the product $\beta\,e^2$
constant and equal to unity, one observes that \rf{action} reduces to the $CP^1$ model (if
$V=0$). Therefore, the configurations \rf{exactclass} are also solutions of the four
dimensional $CP^1$ model.  The ideas of \cite{CPNvortex} were then used to construct multi
vortex solutions for the the four dimensional $CP^N$ model \cite{fkz1,fkz2}. 
Aproximate vortex solutions for the pure Skyrme-Faddeev model, without the potential and $\beta$ terms in  \rf{action}, were constructed in \cite{kundu}.

The static energy density (${\cal H}_{{\rm static}}=-{\cal L}$) associated to \rf{action}
is positive definite if $V>0$, $M^2>0$, $e^2>0$ and $\beta<0$. That is the sector explored
in \cite{glad} and where Hopf soliton solutions were first constructed (for $V=0$). In
addition, that is also the sector explored in \cite{Sawado:2005wa} but with additional
terms involving second derivatives of the ${\vec n}$ field, and where Hopf solitons were
also constructed. The static energy density of \rf{action} is also positive definite for
$V>0$ if
\begin{eqnarray}
M^2>0\,; \qquad  e^2<0\, ; \qquad \beta <0 \, ; \qquad  \beta\, e^2\geq 1
\lab{nicesector}
\end{eqnarray}
That is the sector that agrees with the signature of the terms in the one loop effective
action calculated in \cite{gies} and that we will consider in this
paper. Static Hopf solitons were constructed in \cite{sawadohopfions,todahopfions} for the
sector \rf{nicesector} (with $V=0$) and their quantum excitations, including comparison
with glueball spectrum, were considered in \cite{quantumhopfions}.  An interesting feature
of the Hopf solitons constructed in \cite{sawadohopfions} is that they shrink in size and
then disappear as $\beta\,e^2\rightarrow 1$, which is exactly the point where the vortex
solution of the class \rf{exactclass} exists.

The aim of the present paper is to investigate if vortex solutions for the model
\rf{action} continue to exist when the condition $\beta\, e^2=1$ is relaxed, and so if
they co-exist with the Hopf solitons in \cite{sawadohopfions}. We also aim at the study of
their properties and stability. The idea is to keep the solutions as close to those of the 
class \rf{exactclass} as possible. In order to do that, we follow the ideas of
\cite{babelon} and implement an ansatz based on the $O(2)$ internal symmetry given by the
transformations $u\rightarrow e^{i\alpha}\,u$, together with three commuting
transformations of the Poincar\'e group given by rotations on the plane $x^1\, x^2$, and
translations in the directions $x^0$ and $x^3$. We impose the field configurations to be
invariant under the diagonal subgroups of the tensor product of the internal $O(2)$ group
with each one of the three commuting one parameter subgroups of the Poincar\'e group. The
resulting ansatz is given by
\begin{eqnarray}
u\equiv f\(\rho\)\,e^{i\,\(n\,\vp+\lambda\, z+k\, \tau\)}
\lab{ansatz}
\end{eqnarray}
where $n$ is an integer for $u$ to be single valued, and $\lambda$, $k$ are real
dimensionless parameters, and where we have used the dimensionless polar coordinates
$\(\rho, \varphi, z, \tau\)$, defined by
\begin{eqnarray}
x^0=c\, t= r_0\,\tau\qquad \;\;
x^1=r_0\,\rho\, \cos \vp\qquad\;\;
x^2=r_0\,\rho\, \sin \vp\qquad\;\;
x^3= r_0\,z
\lab{coord}
\end{eqnarray}
and where we have introduced   a length scale $r_0$ given by 
\begin{eqnarray}
r_0^2=-\frac{4}{M^2\,e^2}
\lab{r0def}
\end{eqnarray}
which is positive since we are dealing with $e^2<0$ (see \rf{nicesector}). Note that when
$\lambda=\pm k$ and $f\sim \rho^{\pm n}$, the configurations \rf{ansatz} are of the type
\rf{exactclass}. The ansatz \rf{ansatz} is in fact a generalization to $(3+1)$ dimensions
of the ansatz used in the Baby Skyrme models \cite{wojtekpotential,joaquinansatz}.

The types of potential we will consider in this part are of the form
\begin{eqnarray}
  V\(n_3\)\equiv \frac{\mu^2}{2}\,\(1+n_3\)^{2-\frac{2}{N}}\,\(1-n_3\)^{2+\frac{2}{N}}
  \lab{potdef}
\end{eqnarray}
where $N$ is a non-vanishing integer, $\mu$ a real coupling constant.  It is interesting
to note that when the integer $n$ of \rf{ansatz} has the same modulus as $N$ of \rf{potdef}, one obtains 
analytical solutions of the  form
\begin{align}
  u(\rho, \varphi, z, \tau)=\Bigl(\frac{\rho}{a}\Bigr)^{N}e^{i[\varepsilon\,N\varphi+k(z+\tau)]}
  \lab{zcsolutionintro}
\end{align}
with $\varepsilon=\pm 1$, and where $a=\mid N\mid\left[\frac{(-e^2)\,(\beta e^2-1)M^4}{4\,\mu^2}\right]^{1/4}$.  
Such exact solutions  are valid for all values of the coupling constants.
In particular, for the case $\beta=0$, \rf{zcsolutionintro} are exact solutions of the theory 
\begin{eqnarray}
{\cal L} = M^2\, \partial_{\mu} {\vec n}\cdot\partial^{\mu} {\vec n}
 -\frac{1}{e^2} \, \(\partial_{\mu}{\vec n} \wedge 
\partial_{\nu}{\vec n}\)^2 - \frac{\mu^2}{2}\,\(1+n_3\)^{2-\frac{2}{N}}\,\(1-n_3\)^{2+\frac{2}{N}}
\lab{sfactionwithpot}
\end{eqnarray}
which is the proper Skyrme-Faddeev model in the presence of a potential. In this case we have $a=\mid N\mid\left[\frac{e^2\, M^4}{4\,\mu^2}\right]^{1/4}$, and $\mu^2 \,e^2>0$.

 In some cases we have been unable to find a numerical solution which is
expected in the analytical set, i.e. \rf{zcsolutionintro}. Reasons for this are presented below. Apart from those
cases, we have checked numerically  the existence of the above solution. The numerical simulations are
performed using a standard technique for a differential equation, the Successive Over
Relaxation (SOR) method. In order to further confirm the accuracy and correctness of the
SOR code, some of the results were reproduced by an independent code using Newton's
method, giving typical differences of the order of less than $10^{-4}$.

The paper is organized as follows.  In the next section we briefly describe the extended
Skyrme-Faddeev model. The equations of motion are also introduced in Sec. \ref{sec:vortex}.  We discuss
the Hamiltonian density of the model in Sec. \ref{sec:hamiltonian}.  The method and the solutions of the
integrable sector of the present model are discussed in Sec. \ref{sec:integrable}.  In Sec. \ref{sec:numerics}, we show the
numerical solutions. 
Sec. \ref{sec:application} is devoted to note briefly potential physical applications of our solutions.
A brief summary is presented in Sec. \ref{sec:summary}.

\section{The model}
\label{sec:vortex}
\setcounter{equation}{0}

In terms of the complex scalar field $u$ introduced in \rf{udef} the Lagrangian
\rf{action} becomes

\begin{eqnarray}
{\cal L}=
4\,M^2\,\frac{\partial_{\mu}u\;\partial^{\mu}u^*}{\(1+\u2\)^2} + 
\frac{8}{e^2}\left[ 
\frac{\(\partial_{\mu}u\)^2\(\partial_{\nu}u^*\)^2}{\(1+\u2\)^4}+
\(\beta\,e^2-1\)\,\frac{\(\partial_{\mu}u\;\partial^{\mu}u^*\)^2}{\(1+\u2\)^4}
\right]- V\(\mid u\mid^2\)
\lab{actionu}
\end{eqnarray}
where we have used the fact that $n_3$ is a functional of $\mid u\mid^2$ only, and so is
the potential. The Euler-Lagrange equations following from \rf{actionu}, or \rf{action},
reads
\begin{eqnarray}
\(1+\u2\)\, \partial^{\mu}{\cal K}_{\mu}-2\,u^{*}\,{\cal K}_{\mu}\,
\partial^{\mu} u=-\frac{u}{4}\,\(1+\u2\)^3\,V^{\prime}
\lab{eqmot}
\end{eqnarray}
where $V^{\prime}=\frac{\partial \,V}{\partial \u2}$, and 
\begin{eqnarray}
{\cal K}_{\mu}\equiv M^2\, \partial_{\mu}u 
+\frac{4}{e^2}\,\frac{ 
\left[\(\partial_{\nu}u\partial^{\nu} u\)
\partial_{\mu}u^{*}+\(\beta\,e^2-1\)\,\(\partial_{\nu}u\,\partial^{\nu}u^{*}\)\,
\partial_{\mu} u\right]}{\(1+\u2\)^2}
\lab{kdef}
\end{eqnarray}
We point out that the theory \rf{actionu} possesses an integrable sector defined by the
condition
\begin{eqnarray}
\(\partial_{\mu}u\)^2=0
\lab{eikonal}
\end{eqnarray}
Such condition was first discovered in the context of the $CP^1$ model using the
generalized zero curvature condition for integrable theories in any dimension \cite{afs1},
and then applied to many models with target space being the sphere $S^2$, or $CP^1$ (see
\cite{afs2} for a review). It leads to an infinite number of local conserved
currents. Indeed, \rf{eikonal} together with the equations of motion \rf{eqmot} imply the
conservation of the infinity of currents given by
\begin{eqnarray}
J_{\mu}^G\equiv {\cal K}_{\mu}\, \frac{\delta G}{\delta u}-{\cal K}_{\mu}^*\, \frac{\delta G}{\delta u^*} 
\lab{infinitecurr}
\end{eqnarray}
where $G$ is any functional of $\mid u\mid^2$ only. For the case where the potential
vanishes, the set of conserved currents is considerably enlarged since $G$ can be an
arbitrary functional of $u$ and $u^*$, but not of their derivatives. If in addition to the
condition \rf{eikonal} one takes $V=0$ and $\beta\,e^2=1$, then the equations of motion
reduce to $\partial^2u=0$. It is in that integrable sector that the solutions
\rf{exactclass} lie, and were studied in \cite{vortexlaf}. For theories defined by
Lagrangians which are functionals of the Skyrme term only (pullback of the area form of
the sphere) the currents of the form \rf{infinitecurr} are Noether currents associated to
the area preserving diffeomorphisms of $S^2$ \cite{razumov}. It is possible to define
conditions weaker than \rf{eikonal} that lead to integrable theories associated to Abelian
subgroups of the group of the area preserving diffeomorphisms \cite{joaquinabelian}.

Substituting the ansatz \rf{ansatz} into the equations of motion \rf{eqmot} we get an
ordinary differential equation for the profile function $f$ as
\begin{eqnarray}
\frac{1}{\rho}\,\partial_{\rho}\left[\rho\,\frac{f^{\prime}}{f}\,R\right]-
\frac{\(1-f^2\)}{\(1+f^2\)}\,S\,\left[\Lambda-\(\frac{f^{\prime}}{f}\)^2\right]=
\frac{r_0^2}{4\,M^2}\(1+f^2\)^2\, \frac{\partial\, V}{\partial\mid u\mid^2}
\lab{eqforf}
\end{eqnarray}
where the primes denote derivatives w.r.t. $\rho$, and where
\begin{align}
  \Lambda&= \lambda^2-k^2+\frac{n^2}{\rho^2}\nonu\\
  S&= 1+\beta\,e^2\,\frac{f^2}{\(1+f^2\)^2}\,
  \left[\Lambda+\(\frac{f^{\prime}}{f}\)^2\right]
  \lab{deflambdasr}\\
  R&= 1+\frac{f^2}{\(1+f^2\)^2}\, \left[\(\beta\,e^2-2\)\,\Lambda+\beta\,e^2\,
    \(\frac{f^{\prime}}{f}\)^2\right]\nonu \lab{eqforf2}
\end{align}
With the choice of potential given in \rf{potdef} we get that
\begin{eqnarray}
  \frac{r_0^2}{4\,M^2}\(1+f^2\)^2\, \frac{\partial\, V}{\partial\mid u\mid^2}=
  2\,\frac{r_0^2\,\mu^2}{M^2}\left[\frac{\(2- \frac{2}{ N}\)
      \(f^{2}\)^{\(1-\frac{2}{ N}\)}-\(2+ \frac{2}{ N}\)\(f^{2}\)^{\(2-\frac{2}{ N}\)}}{\(1+f^2\)^{3}}\right]
\end{eqnarray}
We look for solutions satisfying the following boundary conditions
\begin{align}
  \vec{n} \rightarrow
  \begin{cases}
    (0,0,-1) & \text{for } \rho \rightarrow 0 \\
    (0,0,+1) & \text{for } \rho \rightarrow \infty
  \end{cases}
\end{align}
which imply that the profile function should satisfy
\begin{eqnarray}
f\rightarrow 0 \qquad  {\rm for} ~~\rho \rightarrow 0 \qquad\qquad {\rm and} \qquad\qquad
f\rightarrow \infty \qquad  {\rm for} ~~\rho \rightarrow \infty
\end{eqnarray}
Let us then assume the following behavior of the profile function 
\begin{align}
  f = 
  \begin{cases}
    \alpha_0\,\rho^{s_1}\,\bigr(1+\alpha_1\,\rho+\alpha_2\,\rho^2\ldots\bigl) &
    \text{for }\rho \rightarrow 0\\
    \beta_0\,\rho^{s_2}\,\bigr(1+\beta_1\,\frac{1}{\rho}+\beta_2\,\frac{1}{\rho^2}\ldots\bigl) &
    \text{for } \rho \rightarrow \infty
  \end{cases}
  \lab{limitbehavior}
\end{align}
where $s_i>0$, $i=1,2$. Substituting that into the equation \rf{eqforf} one gets the
behavior for small $\rho$ implies that
\begin{eqnarray}
  s_1^2=n^2 
\end{eqnarray}
where $n$ is the integer in the ansatz \rf{ansatz}. The behavior of \rf{eqforf} for large
$\rho$ implies that the relation between $n^2$ and $s_2^2$ depends upon the form of the
potential, and
\begin{align}
  \lambda^2-k^2 =
  \begin{cases}
    \begin{alignedat}{2}
      &8\,\frac{r_0^2\,\mu^2}{M^2} &\qquad& \text{for $N=-1$}
      \\
      &-2\,\frac{r_0^2\,\mu^2}{M^2} && \text{\rm for $N=-2$}
      \\
      &0 && \text{for all other $N$}
    \end{alignedat}
  \end{cases}
\end{align}
where $N$ is the integer appearing in the potential \rf{potdef}, and $\lambda$, $k$ are
the parameters of the ansatz \rf{ansatz}.  Therefore, except for the cases $N=-1$ and
$N=-2$, the waves along the vortex have to travel with the speed of light since the
dependency upon $x^3$ and $x^0$ has to be of the form $x^3\pm c\,t$.  For the dimensionfull
constants $L:=r_0\lambda$, and $K:=r_0k$, the velocity is defined as
\begin{align}
  \frac{Kc}{L}=
  \begin{cases}
    \frac{Kc}{\sqrt{K^2+8\mu^2/M^2}}<c \quad & \text{for $N=-1$} \\
  \frac{Kc}{\sqrt{K^2-2\mu^2/M^2}}>c \quad & \text{for $N=-2$}
  \end{cases}
\end{align}
Therefore, the mode is tachyonic for $N=-2$. $N=-1$ is not tachyonic, but the energy
diverges from the boundary behavior of the potential. In the following analysis, we will
concentrate on the analysis for $N\geqq 1$ (thus $\lambda^2=k^2$).

\section{The Energy}
\label{sec:hamiltonian}
\setcounter{equation}{0}

The Hamiltonian density associated to \rf{actionu} is not positive definite due to the
quartic terms in time derivatives. We shall arrange the Legendre transform of each term in
\rf{actionu} to make explicit such non positive contributions, and write the Hamiltonian
density as (see \cite{bonfim} for details)
\begin{align}
  \begin{split}
  {\cal H} &= 4\, M^2\, \frac{\left[\mid{\dot u}\mid^2+{\vec \nabla}u\cdot
      {\vec\nabla}u^*\right]}{\(1+\u2\)^2} -\frac{24}{e^2}\,\frac{\({\vec \nabla}u\)^2\,\(
    {\vec \nabla}u^*\)^2}{\(1+\u2\)^4} \,\left[ \(\frac{2}{3}\)^2-F^2 \right]
  \\
  &-24\,\frac{\(\beta\,e^2-1\)}{e^2}\,\frac{\left[\mid{\dot u}\mid^2+ \frac{1}{3}\,{\vec
        \nabla}u\cdot {\vec\nabla}u^*\right] \left[ {\vec \nabla}u\cdot
      {\vec\nabla}u^*-\mid{\dot u}\mid^2\right]}{\(1+\u2\)^4} +V\(\mid u\mid^2\)
  \lab{energy}
  \end{split}
\end{align}
where ${\dot u}$ denotes the $x^0$-derivative of $u$, and ${\vec \nabla}u$ its spatial
gradient, and where we have denoted
\begin{eqnarray}
  \frac{{\dot u}^2}{\({\vec \nabla}u\)^2}\equiv \frac{1}{3} + F\, e^{i\,\Phi}
  \lab{fphidef}
\end{eqnarray}
with $F>0$ and $0\leq\Phi\leq 2\pi$, being functions of the space-time coordinates.  Note
that ${\cal H}$ given in \rf{energy} is positive definite for static configurations and
for the range of parameters given in \rf{nicesector}.

Using the ansatz \rf{ansatz} and the coordinates \rf{coord} we get ${\dot
  u}=i\,k\,u/r_0$. The metric on the spatial sub-manifold is given by
$ds^2=r_0^2\(d\rho^2+ \rho^2\,d\varphi^2+dz^2\)$, and so
\begin{eqnarray}
  \({\vec \nabla}u\)^2=\frac{u^2}{r_0^2}\,\(\Omega_{-}-\lambda^2\)\qquad\qquad 
  {\vec \nabla}u\cdot {\vec \nabla}u^*=\frac{f^2}{r_0^2}\,\(\Omega_{+}+\lambda^2\)
\end{eqnarray}
where
\begin{eqnarray}
  \Omega_{\pm}=\(\frac{f^{\prime}}{f}\)^2\pm\frac{n^2}{\rho^2}
  \lab{omegadef}
\end{eqnarray}
In addition one gets that $\frac{{\dot u}^2}{\({\vec \nabla}u\)^2}=\frac{1}{3}+F\,
e^{i\,\Phi}=-\frac{k^2}{\(\Omega_{-}\lambda^2\)}$, and since it is real it follows that
$\Phi=0$ or $\pi$. Therefore, $ \(\frac{2}{3}\)^2 -F^2
=\(\Omega_{-}-\lambda^2-3k^2\)\(\Omega_{-}+k^2-\lambda^2\)/3\(\Omega_{-}-\lambda^2\)^2$. So,
the Hamiltonian density \rf{energy} becomes
\begin{align}
  \lab{energy2}
  \begin{split}
    {\cal H} &= \frac{4}{r_0^4}\,\biggl[M^2\,r_0^2\, \frac{f^2}{\(1+f^2\)^2}\,
      \(\Omega_{+}+\lambda^2+k^2\) \\
      &+2\,\frac{f^4}{\(1+f^2\)^4}
        \biggl\{-
          \frac{1}{e^2}\(\Omega_{-}-\lambda^2-3k^2\)\(\Omega_{-}+k^2-\lambda^2\)
          \\
          &-\frac{\(\beta\,e^2-1\)}{e^2}\,
          \(\Omega_{+}+\lambda^2+3k^2\)\(\Omega_{+}+\lambda^2-k^2\) +\mu^2\,r_0^4\,
          \(f^2\)^{-\frac{2}{N}} \biggr\}
      \biggr]
  \end{split}
\end{align}

\section{The integrable sector}
\label{sec:integrable}
\setcounter{equation}{0}

It is interesting to note that \rf{eqforf}-\rf{eqforf2} with a special choice of the potential \rf{potdef}
have an analytical solution for each topological charge $n$. In fact, solutions of the
integrable equation \rf{eikonal} also become the solutions of the present model. For
$\lambda^2=k^2$, the solutions can be written of the form
\begin{align}
  u(\rho, \varphi, z, \tau)=\Bigl(\frac{\rho}{a}\Bigr)^{n}e^{i[\varepsilon\,n\varphi+k(z+\tau)]}
  \lab{zcsolution}
\end{align}
where $\varepsilon=\pm 1$, and $a$ is a dimensionless  constant to be fixed by the equations of motion. 
Substituting this into the equation \rf{eqforf}-\rf{eqforf2}, we get
\begin{eqnarray}
  (\beta e^2-1)\frac{4n^3}{a^4}\Bigl\{1+\Bigl(\frac{\rho}{a}\Bigr)^{2n}\Bigr\}^{-3}\Bigl[(n-1)
  \Bigl(\frac{\rho}{a}\Bigr)^{2n-4}-(n+1)\Bigl(\frac{\rho}{a}\Bigr)^{4n-4}\Bigr] \nonumber \\
  =\frac{2r_0^2\mu^2}{M^2}\Bigl\{1+\Bigl(\frac{\rho}{a}\Bigr)^{2N}\Bigr\}^{-3}
  \Bigl[(2-\frac{2}{N})\Bigl(\frac{\rho}{a}\Bigr)^{2N-4}
  -(2+\frac{2}{N})\Bigl(\frac{\rho}{a}\Bigr)^{4N-4}\Bigr]
\end{eqnarray}
The constant $a$ determines the scale of the vortex and the equation is satisfied if $n=N$
and
\begin{eqnarray}
  a=\mid n\mid\left[\frac{M^2(\beta e^2-1)}{r_0^2\mu^2}\right]^{1/4}=
  \mid n\mid\left[\frac{(-e^2)\,(\beta e^2-1)M^4}{4\,\mu^2}\right]^{1/4}
  \lab{conditiona}
\end{eqnarray}
Thus, for all possible values of $\beta e^2$ we have  analytical solutions. All those solutions satisfy 
the condition \rf{eikonal}. Clearly the class of solutions contain the special solution at $\beta e^2=1$ found previously in
\cite{vortexlaf} if we take a proper limit of vanishing  potential, i.e.  $\beta e^2\rightarrow 1$ and $\mu^2\rightarrow 0$, with $\frac{\beta e^2-1}{\mu^2}={\rm constant}$. Also, apparently we
have no solution at $\beta e^2 \neq 1$ without any potential because the scale
\rf{conditiona} goes to infinity. Note that the case $\beta =0$ is particularly interesting since it corresponds to the proper Skyrme-Faddeev model (without the extra quartic term) in the presence of a potential. Therefore, the configurations \rf{zcsolution} are exact solutions of  the theory \rf{sfactionwithpot} (for $n=N$),  with $a=\mid N\mid\left[\frac{e^2\, M^4}{4\,\mu^2}\right]^{1/4}$, and $\mu^2 \,e^2>0$.

As we mentioned in Sec.\ref{sec:vortex},  for the sector satisfying  \rf{eikonal}, the model possesses the 
infinite set of conserved currents \rf{infinitecurr}.  In particular, choosing a form of $G=-4i(1+|u|^2)^{-1}$, one gets of the Noether current for
the symmetry of an arbitrary angle $\alpha$, i.e., $u \to e^{i\alpha}u$
\begin{eqnarray}
J_\mu=-4iM^2\frac{u\partial_\mu u^*-u^*\partial_\mu u}{(1+|u|^2)^2}
-i\frac{8}{e^2}(\beta e^2-1)\frac{2(\partial_\nu u\partial^\nu u^*)(\partial_\mu u^*u-u^*\partial_\mu u)}{(1+|u|^2)^4}
\lab{noether}
\end{eqnarray}
For the solution \rf{zcsolution}, we can evaluate the charge per unit length for the solution
\begin{eqnarray}
Q=\int dx_1dx_2 J_0=-8\pi M^2 ka^2r_0
\Bigl[I(n)+\frac{n}{6}\frac{1}{a^2}(\beta e^2-1)\Bigr]
\end{eqnarray}
where $I(n)=\frac{1}{n}\Gamma(1+\frac{1}{n})\Gamma(1-\frac{1}{n})$, with $\Gamma$ being
the Euler's Gamma function.  Here we used an integral formula \cite{gradshteyn}
\begin{eqnarray}
  \int^\infty_0 \frac{x^{\mu-1}dx}{(p+qx^\nu)^{m+1}}
  =\frac{1}{\nu p^{m+1}}\Bigl(\frac{p}{q}\Bigr)^{\frac{\mu}{\nu}}
  \frac{\Gamma(\frac{\mu}{\nu})\Gamma(1+m-\frac{\mu}{\nu})}{\Gamma(1+m)}
\end{eqnarray}

For the Hamiltonian \rf{energy2}, we perform the similar computations.  As a result, we
get the energy per unit length by the integration on the $x_1x_2$ plane.  For $n=1$, the
energy of the static vortex is (in units of $4M^2$)
\begin{eqnarray}
  E_{static}=2\pi+\frac{4\pi}{3}\frac{1}{a^2}(\beta e^2-1)
\label{stt_energy_ana}
\end{eqnarray}
and for $n\geq 2$ they are
\begin{eqnarray}
  E_{static}=2\pi n+\frac{2\pi}{3}\frac{1}{a^2}(\beta e^2-1)(n^2-1) I(n)
\label{stt_energy_ana2}
\end{eqnarray}
Note that the first term is proportional to the topological charge.  The energy per unit
of length of the time-dependent vortex diverges for $n=1$, and for $n\geqq 2$ we obtain
\begin{eqnarray}
  &&E_{wave}
  =2\pi n+\frac{2\pi}{3}\frac{1}{a^2}(\beta e^2-1)(n^2-1) I(n)
  +k^2\Bigl[2\pi a^2I(n)+\frac{2\pi}{3}(\beta e^2-1)n\Bigr]
\label{tt_energy_ana}
\end{eqnarray}
For the limit of $\beta e^2\to 1, \mu^2\to 0$ with keeping $a^2=n^2\sqrt{M^2(\beta
  e^2-1)/r_0^2\mu^2}$ finite, we obtain the energy per unit of length found previously in
\cite{vortexlaf}.  The energy monotonically grows as $k^2$ increases.  Interestingly, the
static vortex has a minimum at $\beta e^2=1.0$ and/or $\mu^2=0.0$ but for the time
dependent vortex there is a minimum of the energy for fixed $\beta e^2$ and $k^2$ and
finite $\mu^2$.  The solutions are confirmed numerically in the subsequent section.

\section{The numerical analysis}
\label{sec:numerics}
\setcounter{equation}{0}

Although the ansatz \rf{ansatz} is given in terms of the polar coordinates, for the
numerical analysis it is more convenient to use a new radial coordinate $y$, defined by
$\rho=\sqrt{\frac{1-y}{y}}$.  Accordingly, we adopt a function $g$ called the profile
function, instead of using $f$, i.e., $f(\rho)=\sqrt{\frac{1-g(y)}{g(y)}}$.

The equation \rf{eqforf} can be promptly rewritten as 
\begin{align}
  &&\frac{d}{dy}\biggl[\frac{y(1-y)}{g(1-g)}g'R\biggr]
  +\Bigl(g-\frac{1}{2}\Bigr)\frac{S}{y(1-y)}\biggl\{\Omega-\biggl(\frac{y(1-y)}{g(1-g)}g'\biggr)^2\biggr\} \nonumber \\
  &&\hspace{2cm}=-\frac{1}{y^2}\frac{r_0^2\mu^2}{M^2}(1-g)^{1-\frac{2}{N}}g^{1+\frac{2}{N}}\Bigl\{4g-2\Bigl(1+\frac{1}{N}\Bigr)\Bigr\}
  \lab{eqforg}
\end{align}
where the primes at this time indicate derivatives w.r.t.$y$ and where
\begin{align}
  \Omega&=(\lambda^2-k^2)\frac{1-y}{y}+n^2 \\
  S&=1+\beta e^2g(1-g)\frac{y}{1-y}\biggl\{\Omega+\biggl(\frac{y(1-y)}{g(1-g)}g'\biggr)^2\biggr\}\\
  R&=1+g(1-g)\frac{y}{1-y}\biggl\{(\beta e^2-2)\Omega+\beta
  e^2\biggl(\frac{y(1-y)}{g(1-g)}g'\biggr)^2\biggr\} \lab{eqforg2}
\end{align}

The energy in the unit of $4M^2$ per unit length for the time-dependent vortex can be estimated 
in terms of following four parts of integrals
of the dimensionless Hamiltonian $H:={\cal H}/4M^2$
\begin{eqnarray}
  E&=&2\pi \int^\infty_0 \rho d\rho H(\rho)=E_2+E_4^{(1)}+(\beta e^2-1)E_4^{(2)}+\frac{r_0^2\mu^2}{M^2}E_0
  \lab{energy_split}
\end{eqnarray}
in which the components are defined as
\begin{eqnarray}
  &&E_2=\pi\int^1_0\frac{dy}{y(1-y)}\biggl\{(k^2+\lambda^2)\frac{1-y}{y}+n^2+\biggl(\frac{y(1-y)}{g(1-g)}g'\biggr)^2\biggr\}g(1-g) 
  \lab{ecomp2}\\
  &&E_4^{(1)}=\pi\int^1_0\frac{dy}{2(1-y)^2}\biggl\{(3k^2+\lambda^2)\frac{1-y}{y}+n^2-\biggl(\frac{y(1-y)}{g(1-g)}g'\biggr)^2\biggr\}\nonumber \\
  &&\hspace{3cm}\times \biggl\{(k^2-\lambda^2)\frac{1-y}{y}+n^2-\biggl(\frac{y(1-y)}{g(1-g)}g'\biggr)^2\biggr\}\bigl(g(1-g)\bigr)^2 
  \lab{ecomp41}\\
  &&E_4^{(2)}=\pi\int^1_0\frac{dy}{2(1-y)^2}\biggl\{(3k^2+\lambda^2)\frac{1-y}{y}+n^2+\biggl(\frac{y(1-y)}{g(1-g)}g'\biggr)^2\biggr\}\nonumber \\
  &&\hspace{3cm}\times \biggl\{(k^2-\lambda^2)\frac{1-y}{y}+n^2+\biggl(\frac{y(1-y)}{g(1-g)}g'\biggr)^2\biggr\}\bigl(g(1-g)\bigr)^2 
  \lab{ecomp42}\\
  &&E_0=2\pi\int^1_0\frac{dy}{y^2} g^{2+\frac{2}{N}}(1-g)^{2-\frac{2}{N}}
  \lab{ecompp}
\end{eqnarray}
For the integrable sector, we should choose $N=n$.

\begin{figure}[t]
\includegraphics[width=8.1cm,clip]{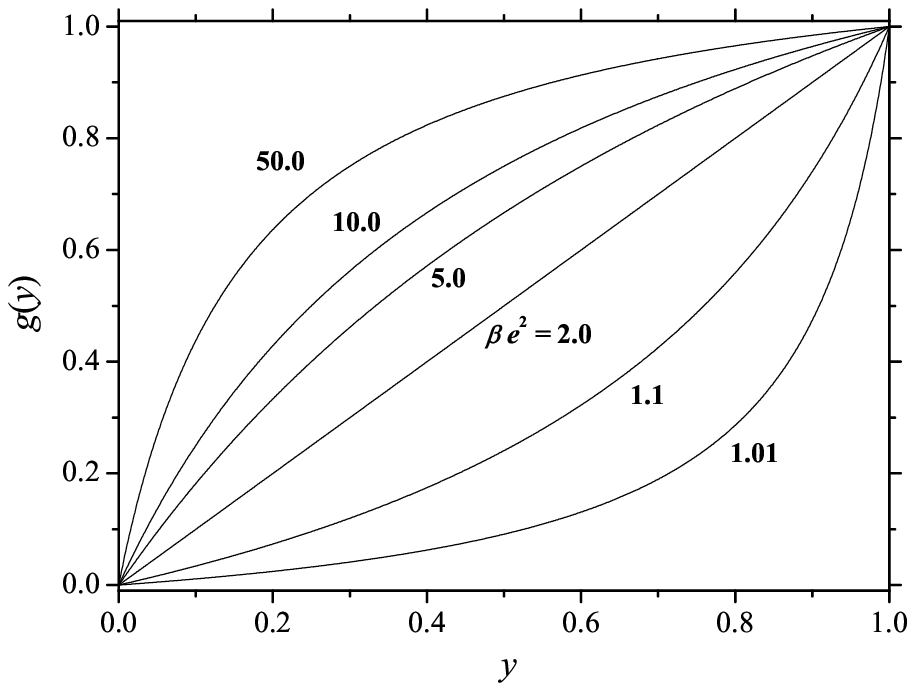}\hspace{-0.5cm}
\includegraphics[width=8cm,clip]{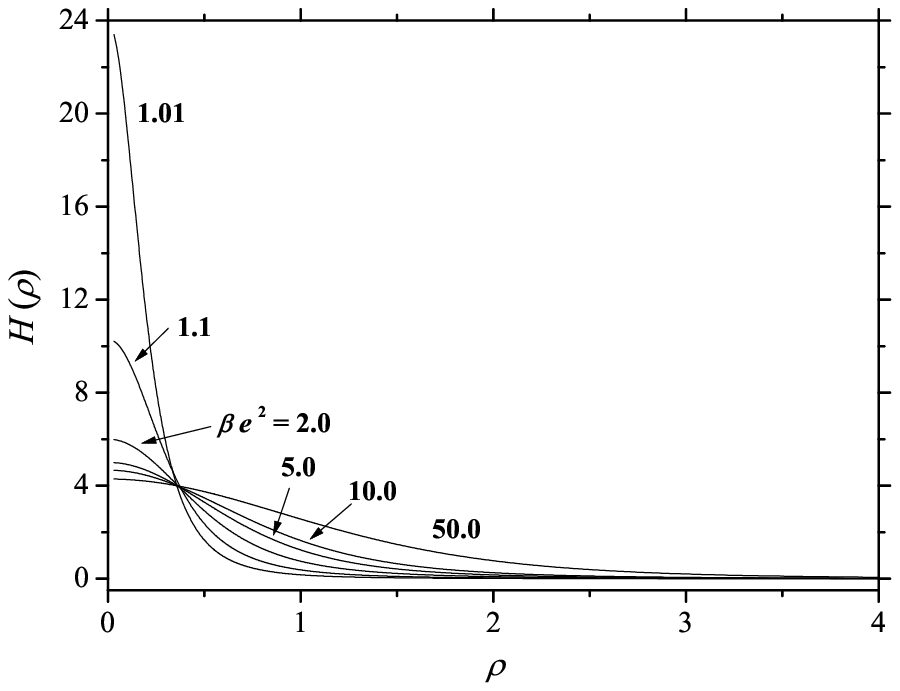}

\caption{\label{profiles1_v04}The $n=1$ profile $g(y)$ and the corresponding Hamiltonian
  density of the real space $H(\rho)$ of $k^2=0.0$ for the constant $r_0^2\mu^2/M^2=1.0$.}
\end{figure}

\begin{figure}
\hspace{3cm}\includegraphics[width=10cm,clip]{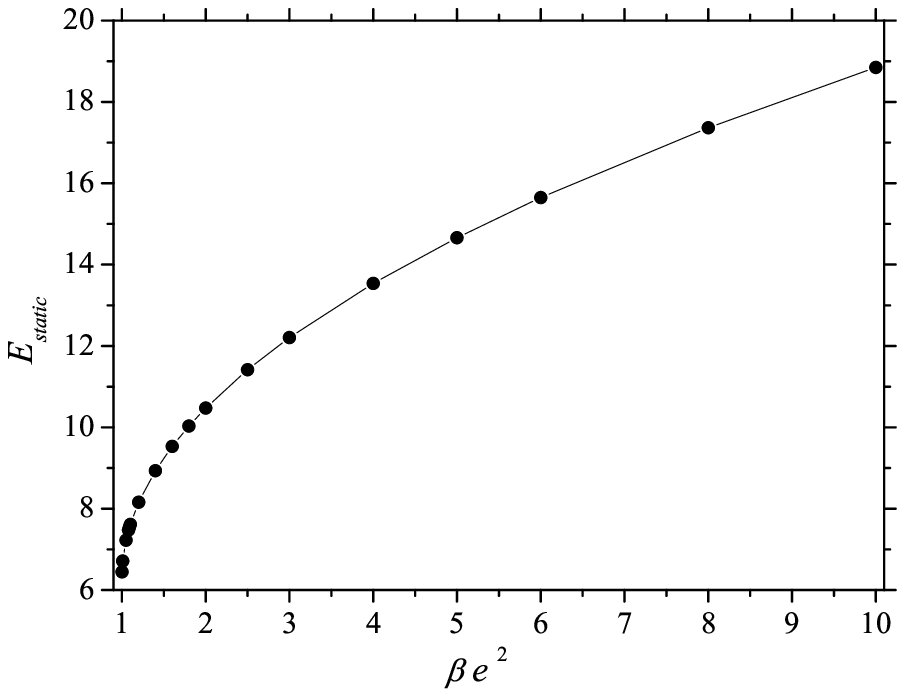}\\
\vspace{-1.0cm}
　\\
\vspace{-2.0cm}
　\\

\includegraphics[width=16cm,clip]{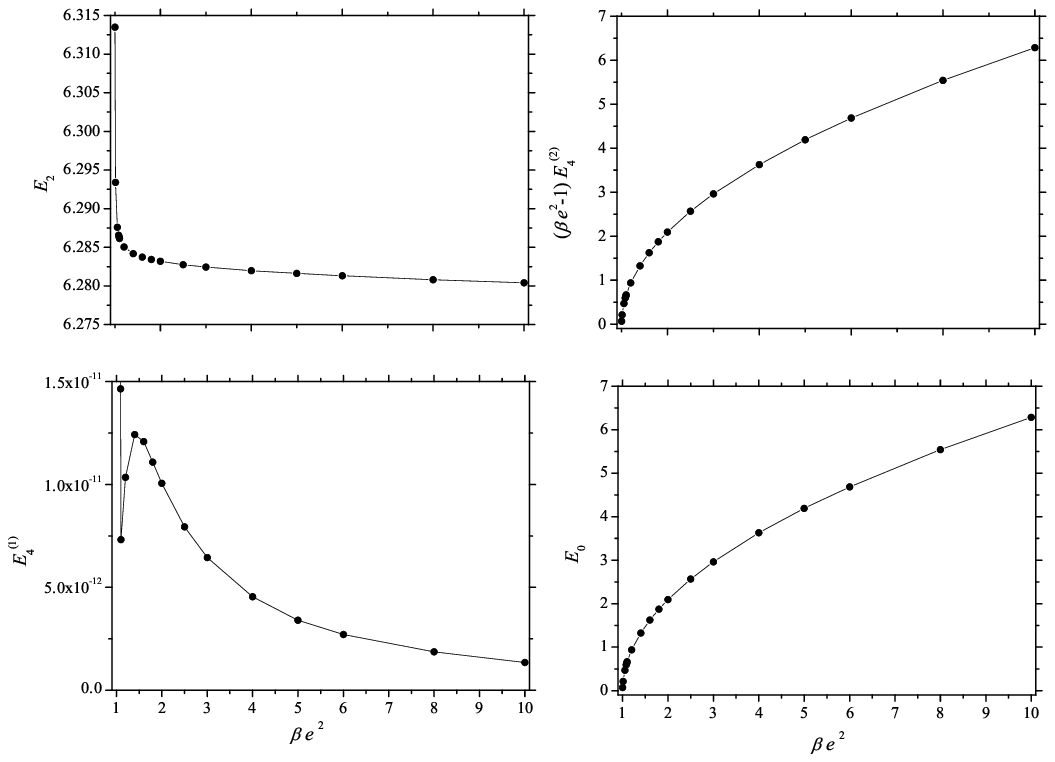}
\caption{\label{energy1_v04}The static energy and its components corresponding to the
  solutions of Fig.\ref{profiles1_v04}.}
\end{figure}

Generally speaking, vortex is an object in three spatial dimensions, thus we have explored
solutions in three spatial dimensions of \rf{action}.  In the three spatial dimensions, the
4th order terms in the Lagrangian (including the Skyrme term) successfully avoid the
non-existence theorem of static and finite energy solutions by Derrick.  However, the
equation \rf{eqforf} of the ansatz \rf{ansatz} is the same as an equation of corresponding
static two spatial dimensions.  This means $z$ component has no essential contribution to
the stability.  In fact, the Derrick's theorem for two spatial dimensions implies that the
contribution to the energy per unit length from quartic terms and the potential must be
equal, namely
\begin{eqnarray}
  E_4^{(1)}+(\beta e^2-1)E_4^{(2)}=\frac{r_0^2\mu^2}{M^2}E_0
\label{derrick}
\end{eqnarray}
Since the solutions of the integrable sector satisfy $E_4^{(1)}=0$, putting together with
$\beta e^2=1$ and $\mu^2=0$, one can confirm that the solution without the potential found
in \cite{vortexlaf} satisfies the above condition automatically. This fact indicates that
the Derrick's argument to the energy per unit length also works well outside of the
integrable sector.

The definition of the scale parameter $a$ given in \rf{conditiona} indicates the existence
of the analytical solution for the same sign of $\beta e^2-1$ and $\mu^2$.  In our
previous study of Hopfions on the extended Skyrme-Faddeev model, we confirmed numerically
the solutions exist only for $\beta e^2>1$ \cite{sawadohopfions}, so we begin our analysis
with the case of $\beta e^2>1$. We shall give comments for the possibility of finding
solution of $\beta e^2<1$ in the next subsection.

\subsection{The solutions of the integrable sector}
\label{subsec:integrable}

The analytic profiles \rf{zcsolution}
can be written in the coordinate $y$ as
\begin{align}
  g(y) = 
  \begin{cases}
    \displaystyle\frac{a^2y}{a^2y+1-y} ~~~~~~~~~~~~ &\text{for ~$n=1$} \\
    \displaystyle\frac{a^4y^2}{a^4y^2+(1-y)^2} ~~~~~~~~~~~&\text{for ~$n=2$}
  \end{cases}
  \lab{anasol}
\end{align}
where $a$ is determined via \rf{conditiona}. 
Apparently \rf{anasol} are solutions of \rf{eqforg}.  Next, we will see whether the
solutions appear or not when we numerically solve \rf{eqforg} without any constraint.
Also, for the obtained solutions we will check the zero curvature condition \rf{eikonal}.

Since \rf{eqforg} is an ordinary second order differential equation, of course there are
several methods to investigate.  However, it is easily noticed that the equation
\rf{eqforg} may exhibit singular-like behavior at the boundary because of the term
$g(1-g)$ of the denominator. Once the computation contains a small numerical error, the
equation quickly diverges.  The numerical method which can safely solve such a difficulty
is well-known, the SOR method. Essentially we have solved the 
following diffusion equation for a field $\tilde{g}(y,t)$
\begin{eqnarray}
  \frac{\partial \tilde{g}}{\partial t}=\omega{\cal A}\Bigl[\tilde{g}, \; \frac{\partial
    \tilde{g}}{\partial y}, \; \frac{\partial^2 \tilde{g}}{\partial y^2}\Bigr]
\end{eqnarray}
in which we employ \rf{eqforg} as ${\cal A}$. Here $\omega$ is called as a relaxation factor
which is usually chosen $\omega= 1.0\sim 2.0$. After a huge number of iteration steps,
the field is relaxed to the static one, i.e, $\tilde{g}(y,t)\to g(y)$, which we are finding.

\begin{figure}[t]
\includegraphics[width=8.1cm,clip]{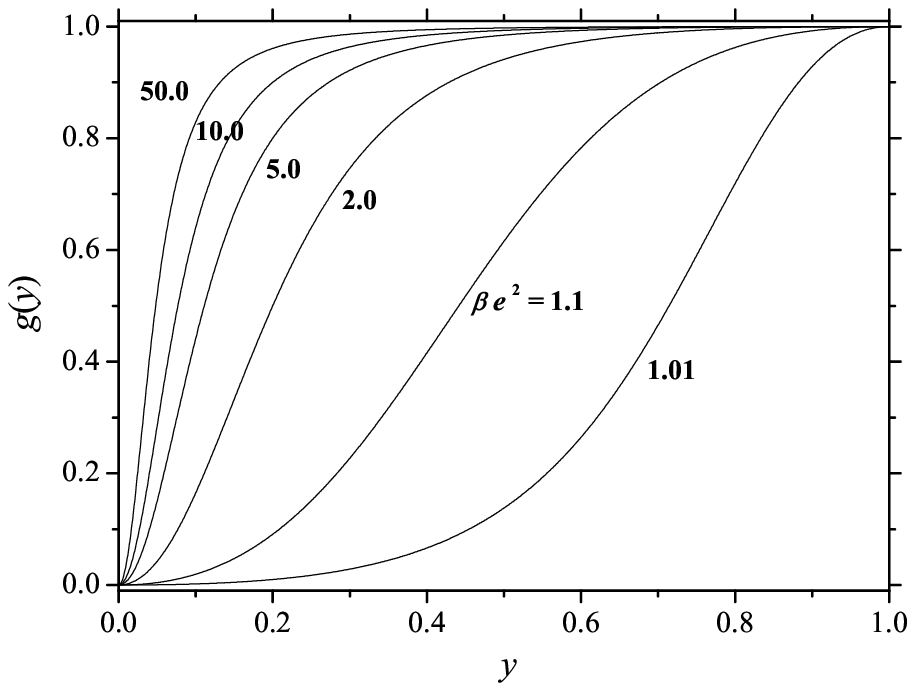}\hspace{-0.5cm}
\includegraphics[width=8cm,clip]{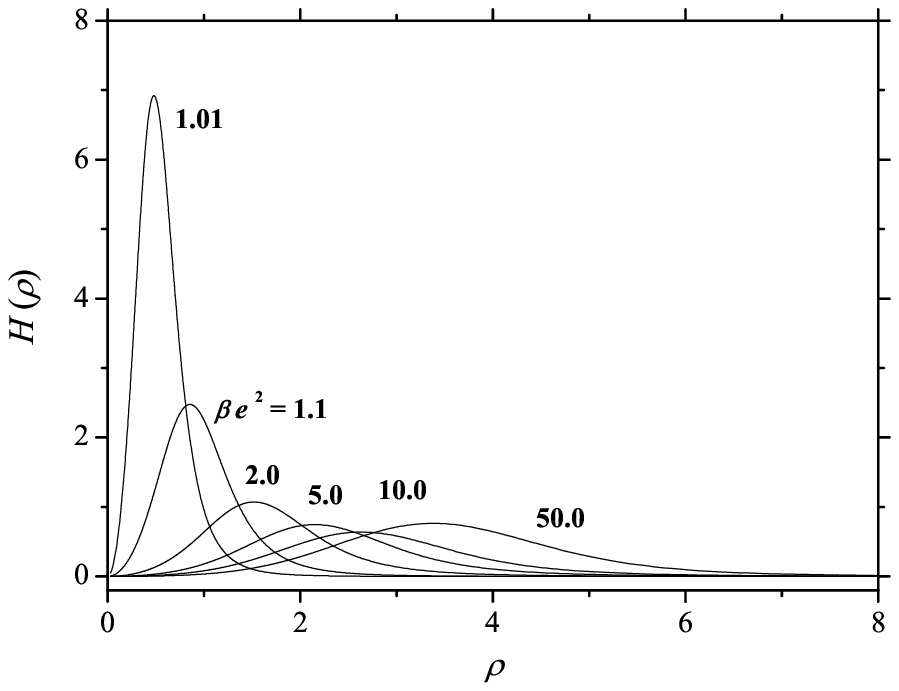}

\caption{
  \label{profiles2_v13}
  The $n=2$ profile $g(y)$ and the corresponding Hamiltonian density of the real space
  $H(\rho)$ of $k^2=0.0$ for the constant $r_0^2\mu^2/M^2=1.0$.}
\end{figure}

\begin{figure}
\hspace{3cm}\includegraphics[width=10cm,clip]{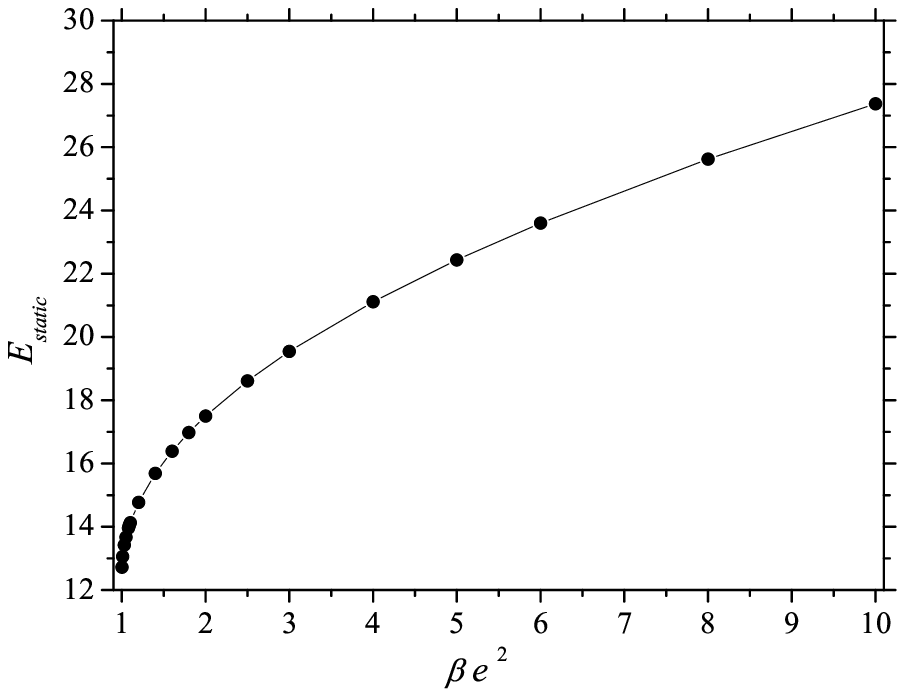}\\
\vspace{-0.0cm}
　\\
\vspace{-2cm}
　\\
\includegraphics[width=16cm,clip]{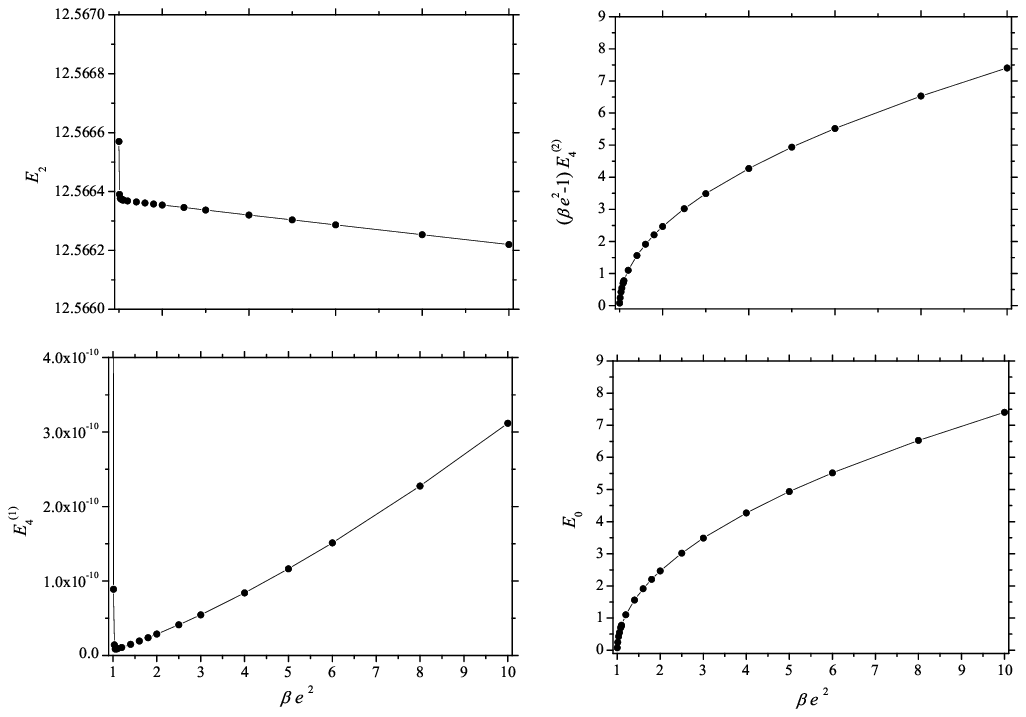}
\caption{
  \label{energy2_v13}
  The static energy and its components corresponding to the solutions of
  Fig.\ref{profiles2_v13}.}
\end{figure}

\begin{figure}[t]
\includegraphics[width=8cm,clip]{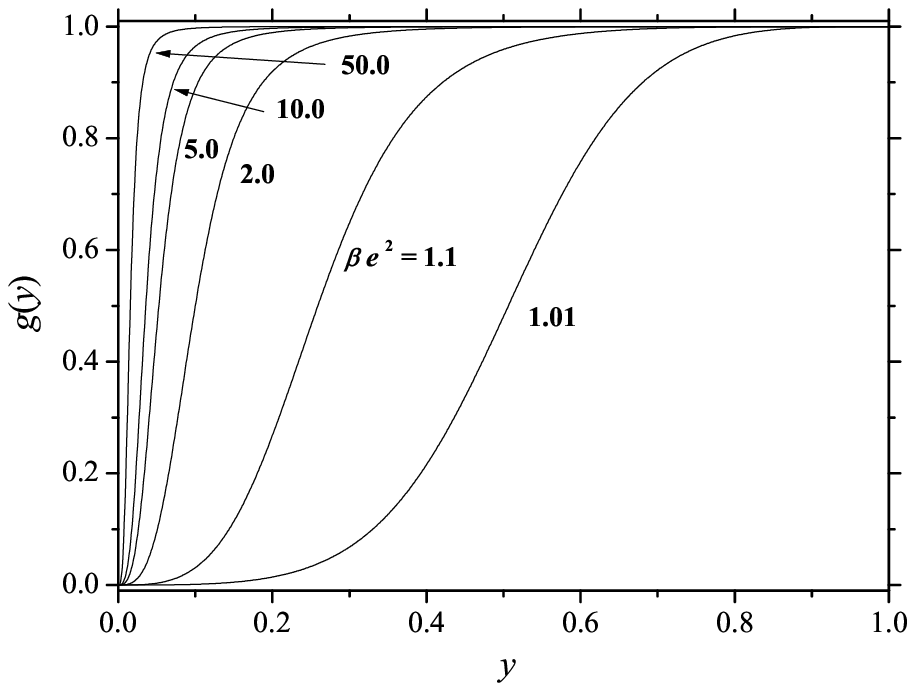}\hspace{-1.cm}
\includegraphics[width=8cm,clip]{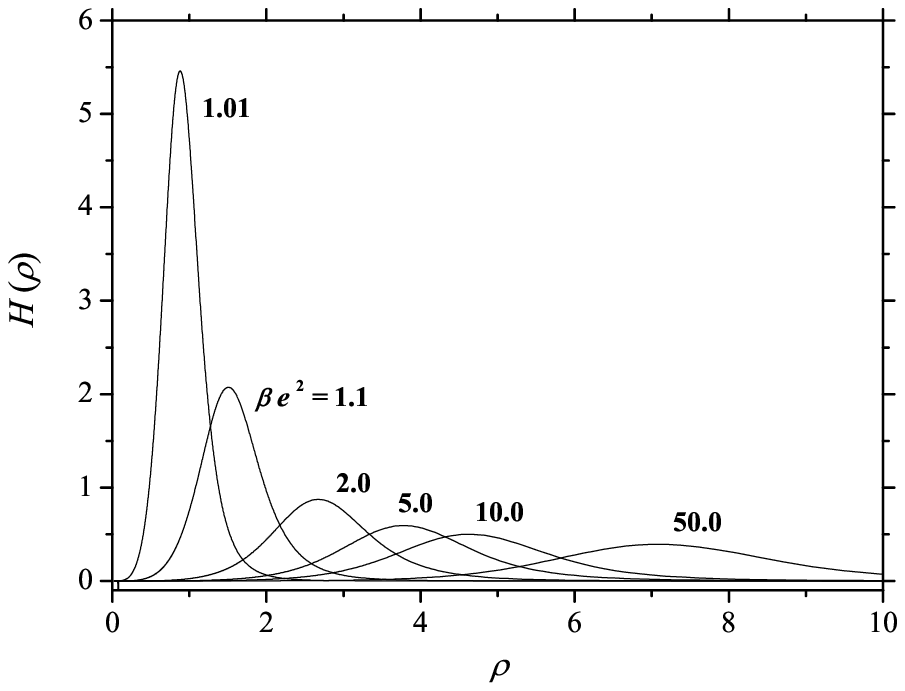}

\caption{
  \label{profiles3_v48}
  The $n=3$ profile $g(y)$ and the corresponding Hamiltonian density of the real space $H(\rho)$
  of $k^2=0.0$. for the constant $r_0^2\mu^2/M^2=1.0$.}
\end{figure}

\begin{figure*}
\hspace{3cm}\includegraphics[width=10cm,clip]{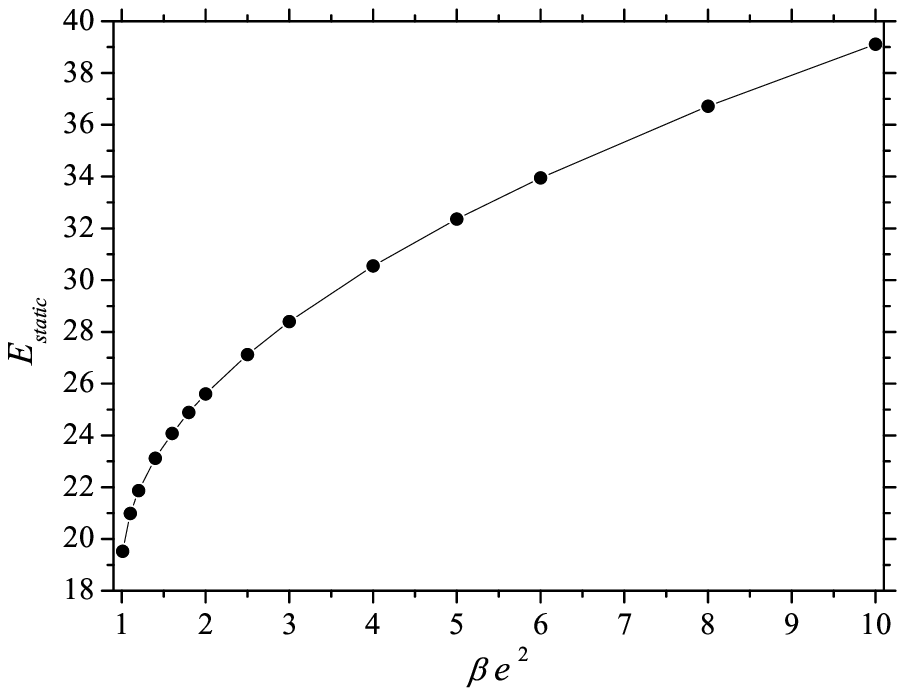}\\
\vspace{-1.0cm}
　\\
\vspace{-1.5cm}
　\\
\includegraphics[width=16cm,clip]{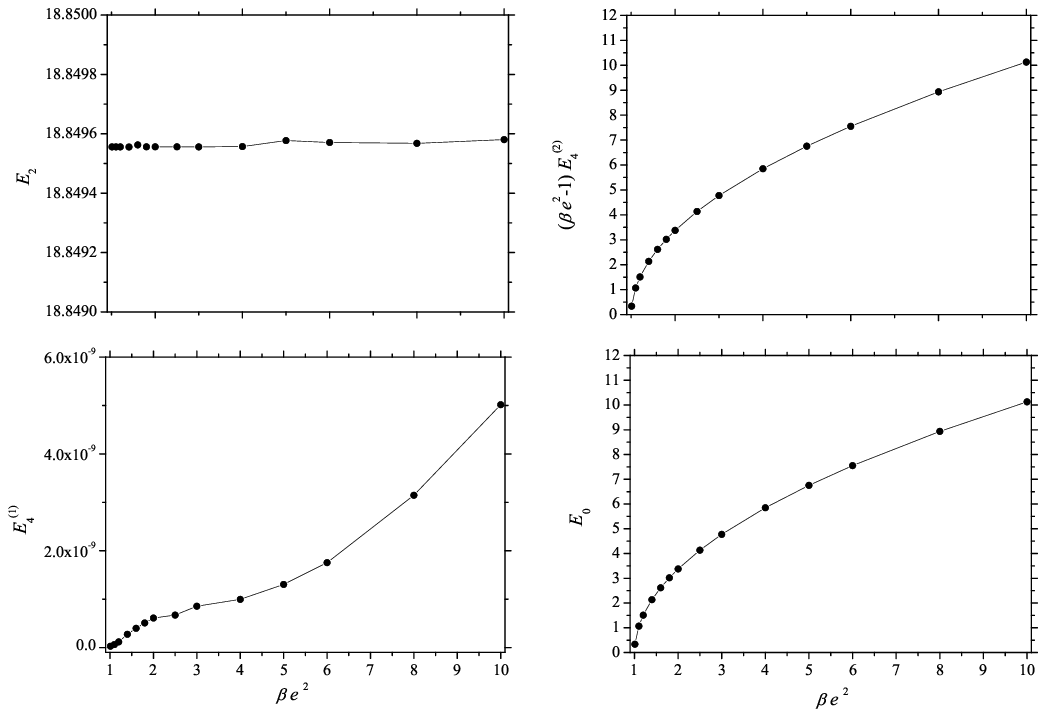}
\caption{
  \label{energy3_v48}
  The static energy and its components corresponding to the solutions of
  Fig.\ref{profiles3_v48}.}
\end{figure*}

\begin{figure*}
\includegraphics[width=8cm,clip]{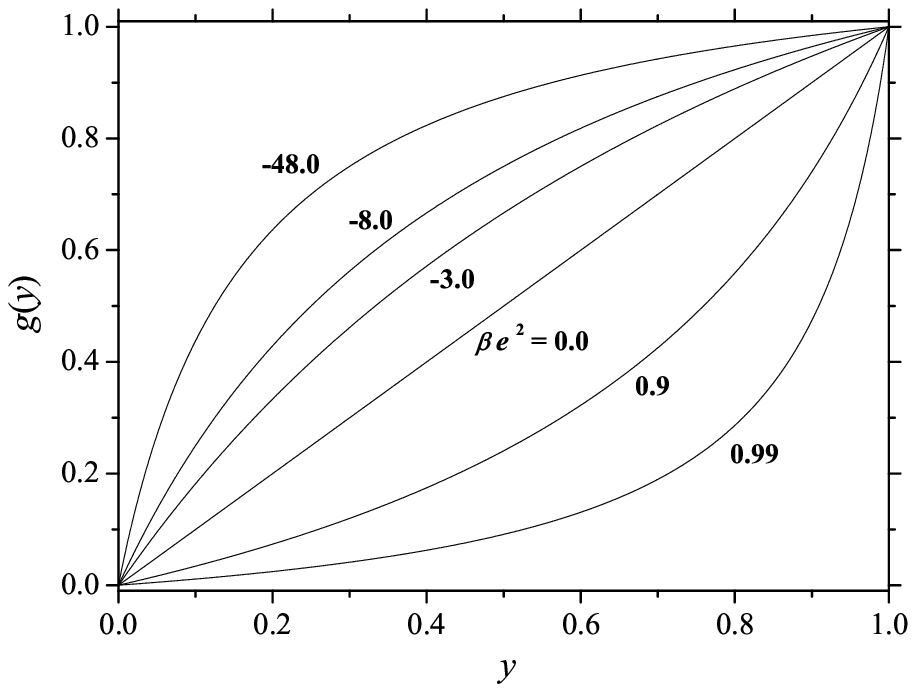}\hspace{-0.5cm}
\includegraphics[width=7.2cm,clip]{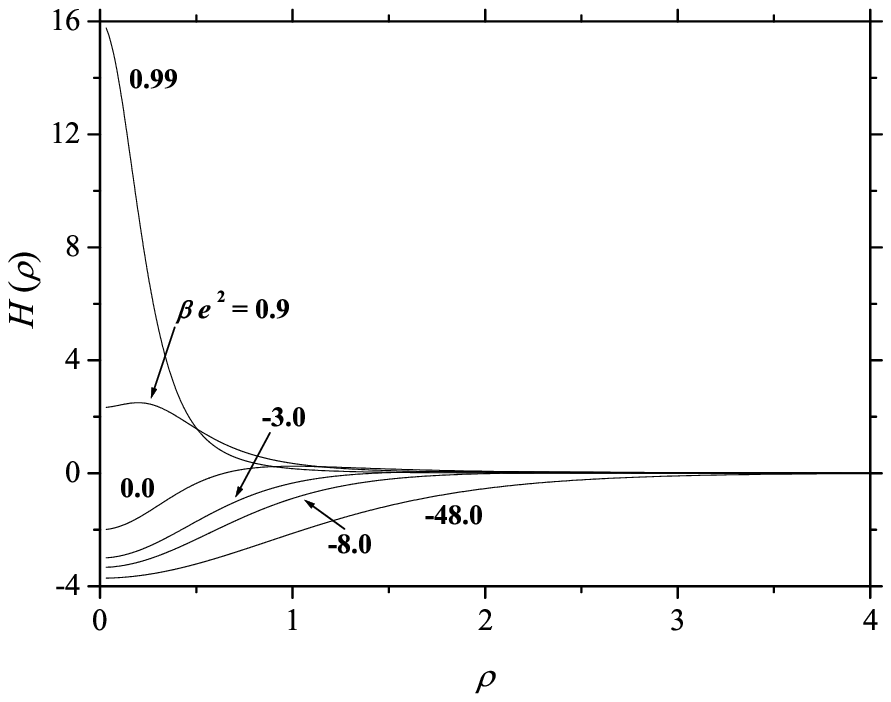}
\caption{
  \label{profile_minus}
  The $n=1$ profile functions and the energy density for the case of $\beta e^2<1$, 
  for the constant $r_0^2\mu^2/M^2=-1.0$.}
\end{figure*}

The first case is $n=1$. From \rf{potdef}, the explicit form of the potential is
\begin{align}
  V_{n=1}=\frac{\mu^2}{2}(1-n_3)^4
\end{align}
In Fig.\ref{profiles1_v04} we present the numerical solution $g(y)$ and the corresponding
Hamiltonian density $H(\rho)$ for $n=1$.  Fig.\ref{energy1_v04} is the energy per unit
length and its components for several values of $\beta e^2$ with fixed $\mu$.  The
function $g(y)$ in Fig.\ref{profiles1_v04} perfectly agrees with the analytical
solution \rf{anasol}.  We shall give a few comments for the components of
the energy.  For the integrable solution, the topological contribution of the energy,
i.e., $E_2$ should be a constant.  Also, $E_4^{(1)}$ is exactly zero for the integrable
solution.  The value of the component $E_4^{(1)}$ in the numerical solution is not exactly zero,
but compatible with zero within the  numerical precision.  Note that the plot seems to blow up for the
vicinity of $\beta e^2=1.0$, but the value is still up to order $\sim 10^{-8}$, so it is
still negligible.  This clearly means that our numerical solutions satisfy the zero
curvature condition and thus belong to the integrable sector.  These numerical errors are
probably originated in the finite number of the mesh points. In a usual case, we used the
number $N_{\rm mesh}=1000$. When we employ a larger number, the value of $E_2$ should be
converge to the constant, i.e., $2\pi$.

For the $n=2$, form of the potential is
\begin{eqnarray}
  V_{n=2}=\frac{\mu^2}{2}(1+n_3)(1-n_3)^3
\end{eqnarray}
thus the potential is zero at both the origin and the infinity.  Fig.\ref{profiles2_v13}
is the profile function and the Hamiltonian density for $n=2$.  
Again the numerical profile and the analytical one \rf{anasol} coalesce. 
Contrary to the case of $n=1$, the density has annular shape .  Fig.\ref{energy2_v13} is the energy per unit
length and its components for several values of $\beta e^2$ and fixed $\mu^2$.  Again we
confirmed that the value of the component $E_1^{(4)}$ is regarded as zero within the
numerical uncertainty.

\begin{figure*}
\centering
\includegraphics[width=12cm,clip]{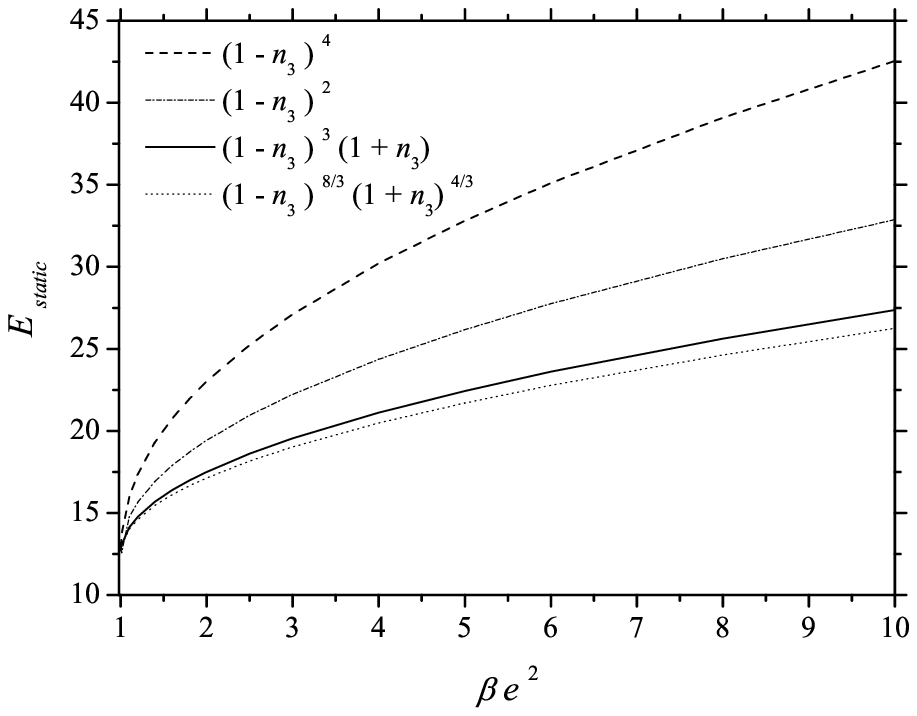}\\
\vspace{-1.0cm}
\includegraphics[width=12cm,clip]{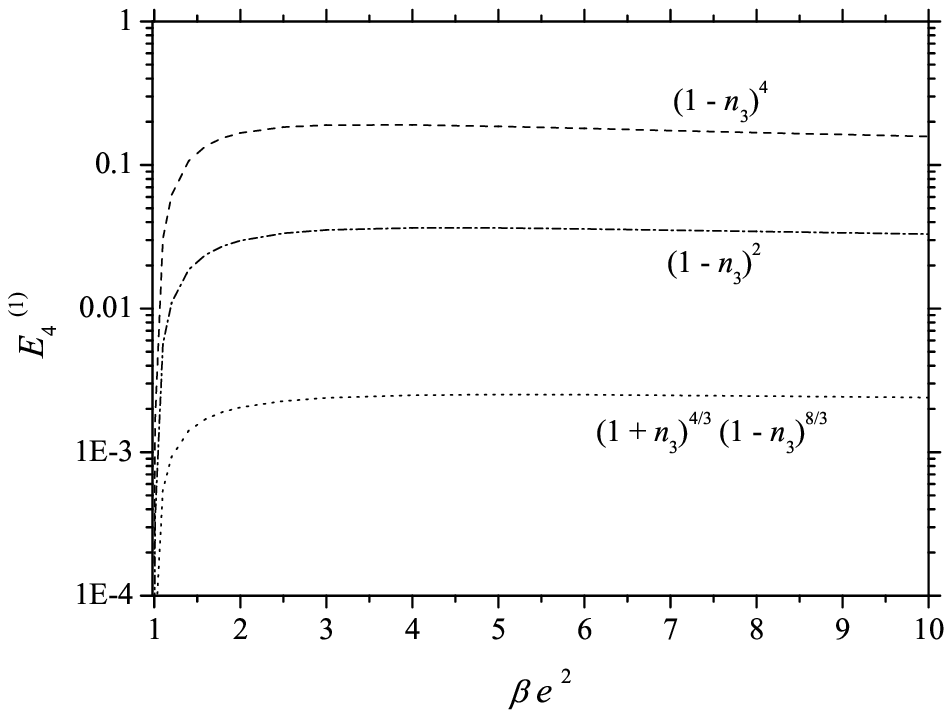}
\caption{
  \label{energy2_pots}
  The $n=2$ static energy and the component $E_4^{(1)}$ for several type of potentials
  $v^i_j$.  $k^2=0$ and $r_0^2\mu^2/M^2=1.0$.}
\end{figure*}

For $n=3$, form of the potential is
\begin{eqnarray}
  V_{n=3}=\frac{\mu^2}{2}(1+n_3)^{4/3}(1-n_3)^{8/3}
\end{eqnarray}
thus again the potential is zero at both the origin and the infinity.
Fig.\ref{profiles3_v48} is the profile function and the Hamiltonian density for $n=3$.  As
is easily seen the radius of the annulus is larger than that of $n=2$.
Fig.\ref{energy3_v48} is the energy per unit length and its components for several values
of $\beta e^2$ and fixed $\mu^2$.  In this case, we face a numerical difficulty. During
the computation by the SOR method, the solution $g$ tends to oscillate around the true value and
sometimes it accidentally goes  below zero at the vicinity of the origin, and then the
computation fails because of the term $g^{1+2/n}$ in \rf{eqforg}.  In order to avoid it,
we employ a finer mesh, i.e., the number is at least $N_{\rm mesh}=3000$.

Until now, we have examined in the case of $\beta e^2>1$.  The formalism leading to \rf{conditiona}
suggests that the choice $\beta e^2<1$ and $\mu^2<0$ might also be possible and the scale
is now defined as
\begin{eqnarray}
  a=|n| \biggl[\frac{M^2(1-\beta e^2)}{-r_0^2\mu^2}\biggr]^{1/4} \quad \text{for $\quad \beta e^2<1,\;\mu^2<0$}
\end{eqnarray}
The result is plotted in Fig.\ref{profile_minus}.  However, existence of such solution seems dubious; 
the energy turns negative at a critical value of $\beta e^2$ thus the solution has no energy lower bound.  
Also, numerically the change of sign of the potential in the equation of motion quickly breaks the computation.

\subsection{The solutions  outside of the integrable sector}
\label{subsec:nonintegrable}

Although we have obtained the analytical solutions for a special form of the potential
\rf{potdef}, we have many options for choice of the potential.  The potentials which we
employed in this paper essentially belong to a class of the generalized Baby-Skyrmion (BS)
potential formally written as
\begin{eqnarray}
  V_{\rm BS}=\frac{\mu^2}{2}v^{\alpha}_\gamma, \; \qquad \qquad v^\alpha_\gamma=(1+n_3)^\alpha(1-n_3)^\gamma
  \label{potential}
\end{eqnarray} 
For $\alpha=0$, the potential is called (the class of) the {\it old}-BS potential which was 
introduced in~\cite{Leese:1989gi}, while $\alpha \neq 0$ is (the class of) the {\it new}-BS
potential~\cite{Kudryavtsev:1997nw}. Our case \rf{potdef} corresponds to
$\alpha=2-2/n,\gamma=2+2/n$.  Note that the possible choice of the potential is certainly
restricted by the analysis of the limiting behavior \rf{limitbehavior} for several
potentials. The results are summarized in Table \ref{table1}.

We can obtain many numerical solutions for the several types of the potentials.  We show
the result of $n=2$ for the potentials $v^0_2,v^0_4,v^{4/3}_{8/3}$; of course these are
not of the form of the analytical solution. Fig.\ref{energy2_pots} presents the energies
and the component $E_4^{(1)}$ for these potentials. For $n=2$, the {\it old}-BS potentials
give higher total energy than the new one. This indicates that the same class of
potentials gives the similar energy and then, for $n=2$ the energy of the {\it new} type
potential $v^{4/3}_{8/3}$ is closest to the integrable sector, which is also plotted in
Fig.~\ref{energy2_pots} for reference.
\begin{table}[t]
\begin{center}
\begin{tabular}{c|ccc}
\hline
$v^\alpha_\gamma$&$n=1$&$n=2$&$n=3$\\
\hline
{\it old}-BS:\; $1-n_3$   & $\times$ &$\times$ & $\times$\\
\hspace{1.6cm}$(1-n_3)^2$ & $\times$ &$\bigcirc$ & $\bigcirc$\\
\hspace{1.6cm}$(1-n_3)^3$ & $\bigcirc$ &$\bigcirc$ & $\bigcirc$\\
\hspace{1.6cm}$(1-n_3)^4$ & $\bigcirc$ &$\bigcirc$ & $\bigcirc$\\\hline
{\it new}-BS:\;$(1+n_3)(1-n_3)$ & $\times$ &$\times$ & $\times$\\
\hspace{1.6cm}$(1+n_3)(1-n_3)^3$ & $\bigcirc$ &$\bigcirc$ & $\bigcirc$\\
\hspace{1.6cm}$(1+n_3)^\frac{4}{3}(1-n_3)^\frac{8}{3}$ & $\bigcirc$ &$\bigcirc$ & $\bigcirc$\\
\hline
\end{tabular}
\end{center}
\caption{The analysis of the limiting behavior of the solutions at both $y=0,1$ for several 
  choice of the potential, and $\bigcirc$ ($\times$) which indicates that there exist (no)
  solutions.}
\label{table1}
\end{table}

\section{Potential physical applications of the solutions}
\label{sec:application}
\setcounter{equation}{0}

Since the model \rf{action} was proposed in the context of Wilsonian 
renormalization group argument of the $SU(2)$ Yang-Mills theory, we expect that the vortex solutions constructed in this paper 
should describe some features of strong coupling regimes, such as the dual superconductor picture
~\cite{Nambu:1974zg}.   
Apart from that, vortices appear in several  areas of physics.  
The Nielsen-Olesen (NO) vortices in the Abelian Higgs model~\cite{Nielsen:1973cs} 
were applied for type II superconductors (SC) and later they have 
extensively been studied in the context of cosmology, i.e., 
the cosmic string~\cite{Hindmarsh:1994re} and the brane-world scenario~\cite{Giovannini:2001hh}. 
The model has a close relationship with the standard electroweak theory, specially when one considers the case of 
a global $SU(2)$ and a local $U(1)$ breaking into a global $U(1)$, where 
the model reduces to an Abelian Higgs model with two charged scalar fields~\cite{Achucarro:1999it}. 
It is interesting to note that the vortices of such model carry the so-called longitudinal electromagnetic 
currents~\cite{Forgacs:2006pm,Volkov:2006ug}. 

\begin{figure*}
\centering
\includegraphics[width=12cm,clip]{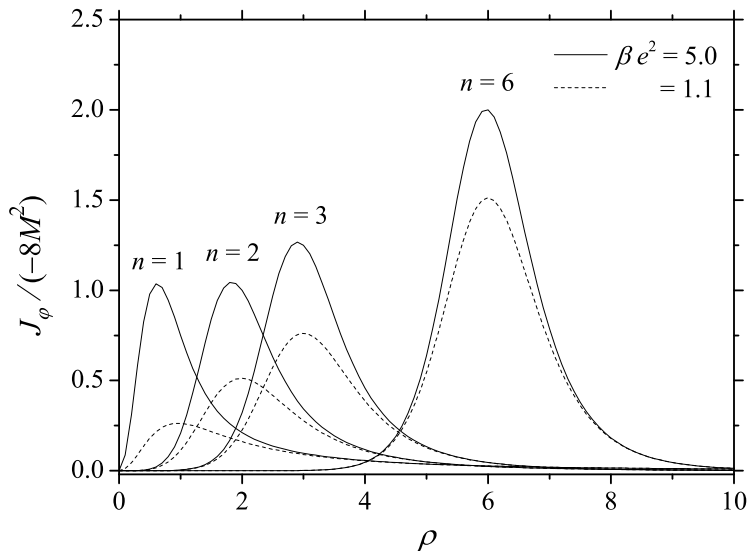}\\
\caption{
  \label{longitudinalcurrent}
The transverse spatial structure of the polar component of the currents (in units of $-1/8M^2$) for the topological charge $n=1,2,3$ and $6$, 
with the parameters $\beta e^2=1.1, 5.0$. We set $r_0^2\mu^2/M^2=1.0$. 
All solutions have a pipe-like structure.  
}
\end{figure*}

In~\rf{noether} we  give the Noether current associated to a global $U(1)$, i.e., $u\to e^{i\alpha}u$, and so  one 
can straightforwardly compute the longitudinal current in the integrable/nonintegrable sector. 
In  Fig.\ref{longitudinalcurrent}, we plot the typical results of the 
transverse spatial structure of the polar component of the current in the case of the integrable sector. 
(Using \rf{zcsolution} and \rf{noether}, 
one can easily see that the radial component of the current is always zero.) 
Note that  for higher winding numbers as well for unit winding number  
the solutions exhibit the pipe-like structure, which was observed in the analysis of \cite{Chernodub:2010sg}.

Our model enjoys a symmetry breaking of the type $O(3)_{\rm global} \to O(2)_{\rm global}$ 
which is similar to $SU(2)_{\rm global}\otimes U(1)_{\rm local}\to U(1)_{\rm global}$.
A notable difference between the NO vortices and ours is that the gauge degrees of freedom are absent in our 
model. If one wishes to discuss the existence of the gauge field in type II SC, however, 
the gauging of the model according to~\cite{Gladikowski:1995sc} should work. 
Confinement or squeezing of the magnetic field into type II SC should be realized in terms of 
the localization of the gauge field into our vortices. 

\section{Summary}
\label{sec:summary}
\setcounter{equation}{0}

We have studied vortex solutions of the extended Skyrme-Faddeev model especially for the
outside of the integrable constraint $\beta e^2=1$. In order to find the solutions, we
introduced potentials of the extension of the Baby-Skyrmion type. We found several
analytical solutions of the model. We also confirmed the existence of the solutions in
terms of the numerical analysis. By using the standard SOR method, we obtained the axially
symmetric solutions for charge up to $n=3$ with several form of the potential for various
value of the model parameters.

In this work, we imposed the axial symmetry to the solution ansatz.  However, solutions
with lower symmetry, such as $\mathbb{Z}_2$-symmetry were found by a numerical simulation
for the Baby-Skyrme model \cite{Hen:2007in}.  It would be interesting to investigate
whether such deformed solutions appear in the extended Skyrme-Faddeev model.  Furthermore,
full 3D simulations of the model will certainly clarify the detailed structure of the
vortices.  The analysis implementing these issues will be discussed in a forthcoming
paper.

\noindent {\bf Acknowledgments} 

We are grateful to the anonymous referee for a careful reading of the manuscript and for valuable
remarks on the physical applications. 
We would like to thank Wojtek Zakrzewski and Pawe\l~Klimas for many useful 
discussions. 
NS would like to thank the kind hospitality at Instituto de F\'isica
de S\~ao Carlos, Universidade de S\~ao Paulo.  He also acknowledges the financial
support of FAPESP (Brazil). JJ would also like to thank the UK Engineering and Physical
Sciences Research Council for support. LAF is partially supported by CNPq-Brazil.


\newpage

\begin{thebibliography}{99}

\bibitem{sf} L.~D.~Faddeev,
  ``Quantization of solitons,'', Princeton preprint IAS Print-75-QS70 (1975); 
\\
in {\em 40 Years in Mathematical Physics} , (World Scientific, 1995).

\bibitem{bp} 
A.~M.~Polyakov and A.~A.~Belavin,
``Metastable States of Two-Dimensional Isotropic Ferromagnets,''  
JETP Lett.\  {\bf 22}, 245 (1975)  [Pisma Zh.\ Eksp.\ Teor.\ Fiz.\  {\bf 22}, 503 (1975)].  


 \bibitem{glad}
  J.~Gladikowski and M.~Hellmund,
  ``Static solitons with non-zero Hopf number,''
  Phys.\ Rev.\  D {\bf 56}, 5194 (1997)
  [arXiv:hep-th/9609035].

\bibitem{solfn}
L.~D.~Faddeev and A.~J.~Niemi,
  ``Knots and particles,''
  Nature {\bf 387}, 58 (1997)
  [arXiv:hep-th/9610193];\\
  ``Toroidal configurations as stable solitons,''
  and arXiv:hep-th/9705176.

\bibitem{sutcliffe}
R.~A.~Battye and P.~M.~Sutcliffe,
  ``Knots as stable soliton solutions in a three-dimensional classical  field
  theory,''
  Phys.\ Rev.\ Lett.\  {\bf 81}, 4798 (1998)
  [arXiv:hep-th/9808129]; 
  ``Solitons, links and knots,''
  Proc.\ Roy.\ Soc.\ Lond.\  {\bf A455}, 4305-4331 (1999) [hep-th/9811077].

\bibitem{hietarinta}
J.~Hietarinta and P.~Salo,
  ``Faddeev-Hopf knots: Dynamics of linked unknots,''
  Phys.\ Lett.\  {\bf B451}, 60-67 (1999) [hep-th/9811053];\\
  ``Ground state in the Faddeev-Skyrme model,''
  Phys.\ Rev.\ D {\bf 62}, 081701(R) (2000).
  
\bibitem{babaev}
  E.~Babaev, L.~D.~Faddeev and A.~J.~Niemi,
  ``Hidden symmetry and knot solitons in a 
charged two-condensate Bose system,''  
Phys.\ Rev.\ B {\bf 65}, 100512 (2002)  
[cond-mat/0106152 [cond-mat.supr-con]].  

  E.~Babaev,
  Phys.\ Rev.\ Lett.\  {\bf 88}, 177002 (2002)
  [arXiv:cond-mat/0106360].
  
 
\bibitem{improve}
  P.~Sutcliffe,
  ``Knots in the Skyrme-Faddeev model,''
  Proc.\ Roy.\ Soc.\ Lond.\  {\bf A463}, 3001-3020 (2007) [arXiv:0705.1468 [hep-th]].\\
  J.~Hietarinta, J.~J\"aykk\"a and P.~Salo,
  ``Relaxation of twisted vortices in the Faddeev-Skyrme model,''
  Phys.\ Lett.\  {\bf A321}, 324-329 (2004) [cond-mat/0309499].\\
  J.~J\"aykk\"a and J.~Hietarinta,
  ``Unwinding in Hopfion vortex bunches,''
  Phys.\ Rev.\  {\bf D79}, 125027 (2009)  [arXiv:0904.1305 [hep-th]].
  
 \bibitem{hietarinta-scatter}
  J.~Hietarinta, J.~Palmu, J.~J\"aykk\"a and P.~Pakkanen,
  ``Scattering of knotted vortices (Hopfions) in the Faddeev-Skyrme model,''
  [arXiv:1108.5551 [hep-th]]. 
  
\bibitem{fn}
L.~D.~Faddeev and A.~J.~Niemi,
  ``Partially dual variables in SU(2) Yang-Mills theory,''
  Phys.\ Rev.\ Lett.\  {\bf 82}, 1624 (1999)
  [arXiv:hep-th/9807069].

\bibitem{chofn}
Y.~M.~Cho,
``A Restricted Gauge Theory,''
Phys.\ Rev.\ D {\bf 21}, 1080 (1980); and 
``Extended Gauge Theory And Its Mass Spectrum,''
Phys.\ Rev.\ D {\bf 23}, 2415 (1981).

\bibitem{wipf}
  L.~Dittmann, T.~Heinzl and A.~Wipf,
  ``A lattice study of the Faddeev-Niemi effective action,''
  Nucl.\ Phys.\ Proc.\ Suppl.\  {\bf 106}, 649 (2002)
  [arXiv:hep-lat/0110026];\\
  ``Effective theories of confinement,''
  Nucl.\ Phys.\ Proc.\ Suppl.\  {\bf 108}, 63 (2002)
  [arXiv:hep-lat/0111037].

\bibitem{newfaddeev}
  L.~D.~Faddeev,
  ``Knots as possible excitations of the quantum Yang-Mills fields,''
  arXiv:0805.1624.\\
  L.~D.~Faddeev and A.~J.~Niemi,
  ``Spin-Charge Separation, Conformal Covariance and the SU(2) Yang-Mills Theory,''
  Nucl.\ Phys.\  {\bf B776}, 38-65 (2007) [hep-th/0608111].

\bibitem{gies}
H.~Gies, 
``Wilsonian effective action for SU(2) Yang-Mills theory with Cho-Faddeev-Niemi-Shabanov decomposition, ''
Phys.\ Rev.\ D {\bf 63}, 125023 (2001), hep-th/0102026.

\bibitem{vortexlaf} 
  L.~A.~Ferreira,
``Exact vortex solutions in an extended Skyrme-Faddeev model, ''
JHEP {\bf 05} (2009)001, 
  arXiv:0809.4303 [hep-th].
  
\bibitem{newsf} L.~A.~Ferreira, P.~Klimas and W.~J.~Zakrzewski, 
``Some properties of (3+1) dimensional vortex solutions in the extended $CP^N$ Skyrme-Faddeev model,''
 arXiv:1111.2338 [hep-th]. 
  
\bibitem{CPNvortex} L. A. Ferreira and P. Klimas, 
``Exact vortex solutions in a $CP^N$ Skyrme-Faddeev type model, ''
JHEP {\bf 10} (2010) 008 [arXiv: 1007.1667].
  
\bibitem{fkz1} L. A. Ferreira, P. Klimas and W.J. Zakrzewski, 
``Some (3+1)-dimensional vortex solutions of the $CP^{N}$ model, ''
Phys. Rev. {\bf D 83} 105018 (2011) [arXiv:1103.0559 [hep-th]].  


\bibitem{fkz2} L.~A.~Ferreira, P.~Klimas and W.~J.~Zakrzewski, 
``Properties of some (3+1)-dimensional vortex solutions of the $CP^{N}$ model, ''
Phys. Rev. {\bf D 84} 085022 (2011) [arXiv:1108.4401 [hep-th]].

\bibitem{kundu} 
  A.~Kundu and Y.~.P.~Rybakov,
  ``Closed vortex type solitons with Hopf index,''
  J.\ Phys.\ A A {\bf 15}, 269 (1982).\\
  C.~-G.~Shi and M.~Hirayama,
  ``Approximate vortex solution of Faddeev model,''
  Int.\ J.\ Mod.\ Phys.\ A {\bf 23}, 1361 (2008)
  [arXiv:0712.4330 [hep-th]].

\bibitem{Sawado:2005wa}
  N.~Sawado, N.~Shiiki and S.~Tanaka,
  ``Hopf soliton solutions from low energy effective action of SU(2)
  Yang-Mills theory,''
  Mod.\ Phys.\ Lett.\  A {\bf 21}, 1189 (2006) [arXiv:hep-ph/0511208].
  
\bibitem{sawadohopfions}
  L.~A.~Ferreira, N.~Sawado and K.~Toda,
  ``Static Hopfions in the extended Skyrme-Faddeev model,''
JHEP {\bf 11}  (2009) 124  [arXiv:0908.3672 [hep-th]].  
  
\bibitem{todahopfions}
  L.~A.~Ferreira, N.~Sawado and K.~Toda,
  ``Axially symmetric soliton solutions in a Skyrme-Faddeev-type model with Gies's extension,''
  J.\ Phys.\ A {\bf A43}, 434014 (2010).  
  
\bibitem{quantumhopfions}  
 L.~A.~Ferreira, S. Kato, N.~Sawado and K.~Toda,
 ``Quantum solitons in the extended Skyrme-Faddeev model,'' 
 Acta Polytechnica {\bf 51}, issue 1, 47 (2011).
 
\bibitem{babelon}
O.~Babelon and L.~A.~Ferreira,
  ``Integrability and conformal symmetry in higher dimensions: A model with
  exact Hopfion solutions,''
  JHEP {\bf 0211}, 020 (2002) [arXiv:hep-th/0210154].
  
  \bibitem{wojtekpotential}
  P.~Eslami, W.~J.~Zakrzewski and M.~Sarbishaei,
  ``Baby Skyrme models for a class of potentials,''
   Nonlinearity {\bf 13} (2000) 1867 [hep-th/0001153].  
  
  \bibitem{joaquinansatz}
  C.~Adam, T.~Romanczukiewicz, J.~Sanchez-Guillen and A.~Wereszczynski,
  ``Investigation of restricted baby Skyrme models,''
  Phys.\ Rev.\  {\bf D81}, 085007 (2010) [arXiv:1002.0851 [hep-th]]. 
  
\bibitem{afs1}
  O.~Alvarez, L.~A.~Ferreira and J.~Sanchez Guillen,
  ``A New approach to integrable theories in any dimension,''
  Nucl.\ Phys.\  B {\bf 529}, 689 (1998)
  [arXiv:hep-th/9710147].
  
  \bibitem{afs2}
  O.~Alvarez, L.~A.~Ferreira and J.~Sanchez-Guillen,
  ``Integrable theories and loop spaces: Fundamentals, applications and new
  developments,''
  Int.\ J.\ Mod.\ Phys.\  A {\bf 24}, 1825 (2009)
  [arXiv:0901.1654 [hep-th]].
  
\bibitem{razumov}
  L.~A.~Ferreira and A.~V.~Razumov,
  ``Hopf solitons and area preserving diffeomorphisms of the sphere,''
  Lett.\ Math.\ Phys.\  {\bf 55}, 143-148 (2001)  [arXiv:hep-th/0012176 [hep-th]].  
  
\bibitem{joaquinabelian}
  C.~Adam, J.~Sanchez-Guillen and A.~Wereszczynski,
  ``Integrability from an Abelian subgroup of the diffeomorphism group,''
  J.\ Math.\ Phys.\  {\bf 47}, 022303 (2006) 
  [hep-th/0511277].  

\bibitem{bonfim}
 L.~A.~Ferreira and A.~C.~Riserio do Bonfim,
  ``Self-dual Hopfions,''  JHEP {\bf 10}, 
119 (2010)  [arXiv:0912.3404 [hep-th]].  
 
\bibitem{gradshteyn}
{\it see e.g.}, 
I.S.Gradshteyn and I.M.Ryzhik, 
"Table of Integrals, Series, and Products,"
(ACADEMIC PRESS, 7 edition, 2007).
  
\bibitem{Leese:1989gi}
  R.~A.~Leese, M.~Peyrard and W.~J.~Zakrzewski,
  ``Soliton scattering in some relativistic models in (2+1)-dimensions,''
  Nonlinearity {\bf 3}, 773 (1990).

\bibitem{Kudryavtsev:1997nw}
  A.~E.~Kudryavtsev, B.~M.~A.~Piette and W.~J.~Zakrzewski,
  ``Skyrmions and domain walls in (2+1)-dimensions,''
  Nonlinearity {\bf 11}, 783 (1998)
  [arXiv:hep-th/9709187].

\bibitem{Nambu:1974zg}
  Y.~Nambu,
  ``Strings, monopoles, and gauge fields,''
  Phys.\ Rev.\  D {\bf 10}, 4262 (1974).\\
  S.~Mandelstam,
  ``Vortices and quark confinement in non-abelian gauge theories,''
  Phys.\ Rept.\  {\bf 23}, 245 (1976).\\
  G.~'t Hooft,
   ``Topology of the gauge condition and new confinement phases in non-abelian
  gauge theories,''
  Nucl.\ Phys.\  B {\bf 190}, 455 (1981).

\bibitem{Nielsen:1973cs}
  H.~B.~Nielsen and P.~Olesen,
  ``Vortex-line models for dual strings,''
  Nucl.\ Phys.\  B {\bf 61} (1973) 45.

\bibitem{Hindmarsh:1994re}
  M.~B.~Hindmarsh and T.~W.~B.~Kibble,
  ``Cosmic strings,''
  Rept.\ Prog.\ Phys.\  {\bf 58}, 477 (1995)
  [arXiv:hep-ph/9411342].

\bibitem{Giovannini:2001hh}
  M.~Giovannini, H.~Meyer and M.~E.~Shaposhnikov,
  ``Warped compactification on Abelian vortex in six dimensions,''
  Nucl.\ Phys.\  B {\bf 619}, 615 (2001)
  [arXiv:hep-th/0104118].

\bibitem{Achucarro:1999it}
  A.~Achucarro and T.~Vachaspati,
  ``Semilocal and electroweak strings,''
  Phys.\ Rept.\  {\bf 327}, 347 (2000)
  [arXiv:hep-ph/9904229].

\bibitem{Forgacs:2006pm}
  P.~Forgacs, S.~Reuillon and M.~S.~Volkov,
  ``Twisted superconducting semilocal strings,''
  Nucl.\ Phys.\  B {\bf 751}, 390 (2006)
  [arXiv:hep-th/0602175].

\bibitem{Volkov:2006ug}
  M.~S.~Volkov,
  ``Superconducting electroweak strings,''
  Phys.\ Lett.\  B {\bf 644}, 203 (2007)
  [arXiv:hep-th/0609112].

\bibitem{Chernodub:2010sg}
  M.~N.~Chernodub and A.~S.~Nedelin,
  ``Pipelike current-carrying vortices in two-component condensates,''
  Phys.\ Rev.\  D {\bf 81}, 125022 (2010)
  [arXiv:1005.3167 [hep-th]].

\bibitem{Gladikowski:1995sc}
  J.~Gladikowski, B.~M.~A.~Piette and B.~J.~Schroers,
  ``Skyrme-Maxwell solitons in (2+1)-dimensions,''
  Phys.\ Rev.\  D {\bf 53}, 844 (1996)
  [arXiv:hep-th/9506099].

\bibitem{Hen:2007in}
  I.~Hen and M.~Karliner,
  ``Rotational symmetry breaking in baby Skyrme models,''
  Nonlinearity {\bf 21}, 399 (2008)
  [arXiv:0710.3939 [hep-th]].






  













\end{thebibliography}
\end{document}